\documentclass[11pt]{article}

\usepackage[a4paper,margin=2.5cm]{geometry}
\usepackage[T1]{fontenc}
\usepackage{lmodern}
\usepackage[utf8]{inputenc}
\usepackage{amsmath,amssymb,amsthm}
\usepackage{graphicx}
\usepackage{epstopdf}   % converts .eps figures on the fly with pdflatex
\graphicspath{{./Figures/}}
\usepackage{multirow}
\usepackage{algorithm}
\usepackage{algorithmic}
\usepackage{caption}
\usepackage{xcolor}
\usepackage[round]{natbib}
\usepackage[colorlinks=true,linkcolor=blue,citecolor=blue,urlcolor=blue]{hyperref}                  % load last

\newcommand{\bA}{\mathbf{A}}
\newcommand{\bB}{\mathbf{B}}
\newcommand{\bI}{\mathbf{I}}
\newcommand{\bL}{\mathbf{L}}
\newcommand{\bP}{\mathbf{P}}
\newcommand{\bQ}{\mathbf{Q}}
\newcommand{\bS}{\mathbf{S}}
\newcommand{\bb}{\mathbf{b}}
\newcommand{\bd}{\mathbf{d}}
\newcommand{\be}{\mathbf{e}}
\newcommand{\bm}{\mathbf{m}}
\newcommand{\bp}{\mathbf{p}}
\newcommand{\bq}{\mathbf{q}}
\newcommand{\bu}{\mathbf{u}}
\newcommand{\bx}{\mathbf{x}}
\newcommand{\blambda}{\boldsymbol{\lambda}}
\DeclareMathOperator*{\minimize}{minimize}

\title{Dual Reverse Time Migration: Efficient Least-squares Imaging with Multiple Scattering in Dual Space}

\author{Mahdi Sonbolestan\textsuperscript{1,*}, Toktam Zand\textsuperscript{1}, Ali Gholami\textsuperscript{1}\\[2mm]
{\small \textsuperscript{1}Institute of Geophysics, Polish Academy of Sciences, Warsaw, Poland}\\
{\small \textsuperscript{*}Corresponding author: \href{mailto:msonbolestan@igf.edu.pl}{msonbolestan@igf.edu.pl}}}

\date{}

\begin{document}

\maketitle

\renewcommand{\abstractname}{Summary} \begin{abstract}
Least-squares reverse time migration (LSRTM) estimates a broadband reflectivity image by iteratively minimizing the data misfit. Despite its superior imaging performance over conventional reverse time migration (RTM), LSRTM is fundamentally limited by the first-order Born approximation and incurs a high computational cost due to the iterative solution of a source-dependent model-space Hessian. We propose dual RTM, a seismic imaging framework that reformulates the inverse problem in the dual space, where the optimization variable is the scattering information of all orders, rather than being limited to the first-order scattering of the Born approximation. Unlike LSRTM, which directly estimates the reflectivity model, dual RTM reconstructs the missing scattering data (dual variable) through a fixed-point iteration governed by the full wave equation, enabling the recovery of higher-order scattering and multiple reflections without repeated background velocity updates. Furthermore, dual RTM replaces the source-dependent model-space Hessian with a source-independent data-space Hessian. A hybrid time--frequency formulation enables this Hessian to be assembled and inverted once in the Fourier domain as an offline preprocessing step, reducing the online computational cost of each dual iteration to that of a single RTM application. Numerical experiments on challenging synthetic benchmarks demonstrate that dual RTM suppresses acquisition footprints, improves amplitude fidelity, and enhances deep imaging where primary reflections alone provide insufficient illumination. These results establish dual RTM as an efficient bridge between conventional RTM/LSRTM and nonlinear waveform inversion for high-resolution large-scale seismic imaging.
\end{abstract}

\noindent\textbf{Key words:} Inverse theory; Numerical approximations and analysis; Waveform inversion; Crustal imaging; Wave scattering and diffraction

\section{Introduction}

Seismic reflection imaging is the principal tool for characterizing the Earth's subsurface, with applications ranging from high-resolution near-surface engineering to complex crustal and subduction-zone studies \citep{Schuster2017,Zand_2024_IAH}. Most modern imaging frameworks are grounded in a linearization of the wave equation relative to a smooth background velocity model. 
This linearization relies on the first-order Born approximation, which simplifies the complex physics of wave propagation by assuming energy interacts only once with subsurface heterogeneities.
Under this assumption, reverse time migration (RTM) \citep{Baysal_1983_RTM, Whitmore_1983_IDM, McMechan_1983_MET} has become the standard for depth imaging due to its ability to handle turning waves, prismatic reflections, and steep dips by solving the two-way wave equation.
It generates reflectivity images by applying the adjoint of the Born modeling operator---specifically, the zero-lag cross-correlation of forward-propagated source and backward-propagated receiver wavefields
\citep{Symes_2008_MVA}.

Nevertheless, RTM approximates the inverse Born operator by its adjoint, effectively replacing the model-space Hessian with the identity matrix. Consequently, the resulting image is a blurred representation of the true reflectivity, suffering from the spatial filtering effects of the acquisition geometry and the limited bandwidth of the source---leading to an uneven illumination and distorted amplitudes \citep{Plessix_2004_FDF}. Least-squares reverse time migration (LSRTM) was developed to mitigate these effects by reformulating imaging as an iterative optimization problem  \citep{Lailly_1983_SPS, Nemeth_1999_LSQ, Dai2012}. By minimizing the misfit between Born-modeled and observed data, LSRTM progressively de-blurs the image and compensates for the Hessian's effects, restoring amplitude fidelity and enhancing spatial resolution \citep{Dai2012, Zand_2023_LSR}. 

However, these benefits come at a prohibitive computational price, as LSRTM requires solving a massive least-squares problem. Minimizing the objective function typically necessitates iterative solvers, such as conjugate gradients \citep{Kelley_1999_IMO}, which require repeated forward Born modeling and adjoint wavefield propagations for every source at every iteration---totaling $3N_{\texttt{iter}}N_s$ wave-equation solves, where $N_{\texttt{iter}}$ is the number of LSRTM iterations and $N_s$ is the number of seismic sources. For large-scale crustal or regional datasets, this linear scaling with the number of iterations often renders the method computationally infeasible. A further practical challenge lies in the selection of a stable step length. While analytical formulas exist, they rely on the assumption of exact adjointness between the Born modeling and migration operators---an assumption frequently violated in practical numerical implementations. Consequently, expensive numerical line-search methods, such as the secant method, are often required. These can demand additional wave-equation solves per iteration just to estimate a single step length, significantly increasing the total online cost of the inversion.

Various strategies have been proposed to mitigate the computational burden of LSRTM, including phase encoding \citep{Romero_2000_PEPM, Dai2012}, preconditioning \citep{Liu_2018_ODL}, wavenumber-domain Hessian approximations \citep{Hu2001, Guitton2004}, memory-saving strategies for adjoint-state
methods \citep{Symes_2007_RTM,Zand_2019_CIR,gholami2019efficient}, and using acceleration approaches such as the Anderson acceleration (AA) method \citep{anderson1965iterative,Yang_2021_AAS,zand-2021}. Notably, one-step data-domain approaches have been developed to reduce the total cost to only two RTM applications \citep{Liu_2018_ODL}. 

However, these methods remain strictly confined within the first-order Born approximation, accounting only for single-scattered energy.
Consequently, the complex multiple-scattered wavefield---comprising internal and surface-related multiples---is typically discarded as noise. Yet, multiples carry independent and often superior illumination compared to primaries; they have different travel paths, exhibit smaller reflection angles, and provide a denser fold \citep{berkhout2017utilization, nath2020imaging}. 
By ignoring this higher-order scattering energy, conventional LSRTM attempts to fit the data with an inadequate physical model, leading to modeling mismatches that manifest as coherent artifacts and amplitude distortions in the final image. However, the correct mapping of multiples is particularly important for enhancing vertical resolution and imaging in structurally complex environments, such as sub-salt or sub-basalt settings, where primaries alone provide insufficient coverage \citep{zhang2014least,Fleury2013, Davydenko2017}. In particular, the contribution of surface-related multiples is essential for recovering shallow interfaces using coarse acquisition geometries \citep{zhang2014least} and for imaging structures beneath the environmental barriers that cause acquisition gaps \citep{nath2020imaging}.

These observations have motivated a growing body of work aimed at incorporating multiple scattering into the imaging process while reducing sensitivity to background model inaccuracies \citep{malcolm2011recursive, nath2020imaging, jiang2007migration, verschuur2015removing}. Particularly, modifications in the multiple imaging based on LSRTM approach, studied as nonlinear extensions of RTM (NLRTM) \citep{Fleury2013}, account for the fundamental nonlinear relationship between model contrasts and scattered data by iteratively re-simulating the wavefield.
Similarly, full wavefield migration (FWM) \citep{Davydenko2017, Davydenko2021FWLSRTM} uses a closed-loop inversion to jointly image primaries and multiples, or higher-order Born modeling is implemented to generate infinite orders of multiple scattering \citep{makinen2018multiples,chen2022internal}, while other frameworks augment the seismic source with the reflected wavefield to improve illumination \citep{zhang2014least, Tu2015}. 
Despite their theoretical merits, these approaches face significant practical hurdles. A primary challenge is the emergence of crosstalk artifacts, where primary and multiple energy are incorrectly mapped to the same spatial locations. Mitigating these artifacts typically requires expensive wavefield separation, precise prior knowledge of reflecting boundaries, or intensive parameter tuning. Another challenge is the prediction of multiples: although modeling them by solving the full wave equation is straightforward, it is not possible within a linearized inversion based on first-order Born modeling. 
Modeling and imaging of the internal multiples are even more challenging than free-surface-related multiples, where at least one of the reflectors, which is the free surface, is known. Studies on internal multiples follow two directions: one-way approaches, such as full wavefield modeling \citep{Berkhout_2014_OFS} and wavefield extrapolation \citep{malcolm2009seismic}, which have limited ability to image steep dips; and two-way approaches, such as higher-order Born modeling \citep{chen2022internal}, which remain computationally demanding.

In this paper, we introduce Dual Reverse Time Migration (DRTM), a framework that unifies the strengths of RTM, LSRTM, and nonlinear waveform inversion. Unlike conventional imaging methods derived from the reduced-space formulation of full-waveform inversion (FWI), DRTM is rooted in the dual formulation of FWI \citep{Aghazade_2025_FAF}. By imposing the wave equation as a hard constraint through Lagrange multipliers, we transform the imaging task into a saddle-point (min--max) optimization \citep{Aghamiry_2019_IWR,Gholami_2022_EFW}. This shift allows the framework to explicitly manage the wavefield and dual variables, moving beyond the limitations of reduced optimization problems that typically rely on the linearized Born approximation \citep{Operto_2023_FWI}.

The physical advantage of DRTM lies in its decision to retain the dual variables throughout the inversion process. By linearizing only with respect to model parameters while keeping the dual variables active, the predicted data are no longer restricted by the linear Born operator. Instead, these dual variables are iteratively updated through solutions of the full wave equation, enabling them to capture the nonlinear data mismatch, including internal and surface-related multiples. This allows DRTM to recover information beyond the Born approximation while maintaining a fixed background velocity model, effectively bypassing the need for repeated, expensive model updates to compensate for higher-order scattering.

Furthermore, DRTM achieves extreme computational efficiency by shifting the Hessian from the model domain to the data domain. Unlike LSRTM, which requires repeated wave-equation solves to iteratively approximately compensate for the model-space Hessian, our dual formulation utilizes a direct inversion of the data-space Hessian \citep{Sonbolestan2025}. This is facilitated by a hybrid time--frequency implementation where receiver Green's functions are computed in the time domain and inverted in the frequency domain as a one-time offline cost. Because this Hessian depends strictly on the receiver geometry, it can be reused for all sources throughout the inversion, keeping the online computational cost equivalent to only two wave-equation solves per source per iteration.

The DRTM formulation is independent of the specific wave physics. It is developed here for the acoustic case, but because the dual iteration acts entirely in the data space, the transition to the multiparameter elastic case requires only the substitution of the wave operator and its sensitivity kernel, leaving the dual-space formulation unchanged; we demonstrate the performance of the method using synthetic acoustic data.

The remainder of this paper is organized as follows. Section~\ref{sec:pre} introduces the linearized Born model, RTM, and LSRTM, together with their modeling and computational limitations. Section~\ref{sec:method} then develops the proposed dual formulation, deriving the dual scattering source, its projection into the data space, the data-space Hessian, and the resulting fixed-point iteration for the dual variables. Section~\ref{sec:complexity} analyses the computational cost of the proposed method and compares it with RTM and LSRTM. 
Section~\ref{sec:results} presents numerical experiments on synthetic benchmarks representing marine exploration and crustal-scale ocean-bottom seismometer (OBS) acquisitions. The experiments include imaging data without surface-related multiples to demonstrate the method's ability to leverage the contribution of internal multiples in imaging and the improvement of its results compared with RTM and LSRTM, which only image primary reflections. Furthermore, each experiment is repeated with data that include surface-related multiples to demonstrate their individual impact on the results.
Section~\ref{sec:conclusion} summarizes the findings and outlines directions for future work. 

%%%%%%%%%%%%%%%%%%%%%%%%%%%%%%%%%%%%%%%%%%%%%%%%%%%%%%%%%%%%%%%%%%%%%%%%%%%%%%%%%%%%%%%%%

\section{Preliminary} \label{sec:pre}
The acoustic wave equation in a medium with variable velocity and constant density is given by
\begin{equation} \label{waveEqc}
m^*(\mathbf{x}) \frac{\partial^2}{\partial t^2}u^*(t,\mathbf{x};\mathbf{x}_s) - \Delta u^*(t,\mathbf{x};\mathbf{x}_s)=b(t,\mathbf{x};\mathbf{x}_s),
\end{equation}
where $m^*(\bx)$ denotes the model parameters (squared slowness), $u^*(t,\bx;\bx_s)$ is the associated seismic wavefield, and $b(t,\bx;\bx_s)$ represents the source term. The variable $t$ denotes time, while $\bx$ denotes the spatial coordinate in two or three dimensions. The source locations are denoted by $\bx_s$, $s=1,2,\ldots,N_s$, where $N_s$ is the total number of sources. Finally, $\Delta$ denotes the Laplacian operator. For simplicity of exposition, the methodology below is developed throughout for the acoustic case; because the dual-space formulation depends on the wave physics only through the wave operators, the extension to other wave operators, such as the elastic case, follows by substituting their instances while the dual-space formulation remains unchanged.

The wavefield is sampled at the receiver locations $\bx_r, r=1,2,\ldots,N_r$, where $N_r$ is the number of receivers, to generate the observed data,
\begin{equation}
d(t,\bx_r;\bx_s)=u^*(t,\bx_r;\bx_s) +e_s,
\end{equation}
where $e_s$ denotes additive noise. These recorded data serve as the input to the imaging algorithms for estimating the model parameters $m(\bx)$.

The forward problem consists of computing the seismic data $\bd^*$ for a given model parameter vector $\bm^*$, where $\bd^*\in\mathbb{R}^{M}$ and $\bm^*\in\mathbb{R}^{N}$ denote the discretized data and model vectors, respectively. This is accomplished by discretizing the computational domain $\Omega \subset \mathbb{R}^d$, with $d=2$ or $3$, thereby representing the model parameters, wavefields, and source term on a numerical grid and replacing the continuous differential operators with their discrete counterparts. The resulting system of equations is then solved using a suitable time-stepping scheme to propagate the wavefield from each source location. For a detailed description of the spatial discretization and time-integration methods used to solve the forward problem, we refer the reader to \citet{alford1974accuracy, igel2017computational,zand2026MNAD}. 

The forward modeling operator can be naturally decomposed into two consecutive steps: (i) solving the discretized wave equation to obtain the wavefield throughout the computational domain, and (ii) sampling the wavefield at the receiver locations to generate the predicted seismic data. After discretization in space and time, the wave equation for the $s$-th source can be written compactly as
\begin{equation}
\label{eq:forward_wave}
\bA(\bm^*)\bu_s^*=\bb_s, \quad s=1,...,N_s,
\end{equation}
where $\bA(\bm^*)\in\mathbb{R}^{\hat{N}\times \hat{N}}$ denotes the discretized wave operator, $\bu_s^*\in\mathbb{R}^{\hat{N}}$ is the discretized space--time wavefield, and $\mathbf{b}_s\in\mathbb{R}^{\hat{N}}$ is the source vector. The dimension $\hat{N}$ equals the product of the number of spatial grid points and the number of time steps.
Once the wavefield has been computed, the predicted data are obtained by sampling the wavefield at the receiver locations,
\begin{equation}
\label{eq:sampling}
\bd_s=\bP\bu_s^*+\be_s,\quad s=1,...,N_s,
\end{equation}
where $\bP\in\mathbb{R}^{M\times \hat{N}}$ is the sampling operator and $M$ is the number of recorded data samples for one source. For simplicity, we assume that all source experiments share the same receiver geometry so that the sampling operator $\mathbf{P}$ is identical for every source.

Combining Eqs. \eqref{eq:forward_wave} and \eqref{eq:sampling} yields the nonlinear forward operator
\begin{equation}
\label{eq:forward_operator}
G_s(\bm) =\bP\bA(\bm)^{-1}\bb_s,\quad s=1,...,N_s,
\end{equation}
which maps the model parameters to the predicted data for the $s$-th source.

The imaging problem is an inverse problem in which the objective is to estimate the subsurface model parameters from the recorded seismic data. Given the observed data for all source experiments, the problem can be stated as
\begin{equation}
\label{eq:inverse_problem}
\text{Given } \{\bd_s\}_{s=1}^{N_s}, \text{ find } \bm\in\mathcal{M}
\text{ such that }
\bd_s \approx G_s(\bm), \qquad s=1,\ldots,N_s,
\end{equation}
where $\mathcal{M}\subseteq\mathbb{R}^N$ denotes the set of admissible models. This is usually cast as an optimization problem \citep{Tarantola_1982_GNI,Pratt_1998_GNF}:
\begin{equation}
\label{eq:FWI}
\minimize_{\bm \in\mathcal{M}} ~J_{\texttt{FWI}}(\bm),\quad\mbox{where}\quad J_{\texttt{FWI}}(\bm):= \frac{1}{2}\sum_{s=1}^{N_s} \|G_s(\bm) - \mathbf{d}_s\|_2^2.
\end{equation}
The inverse problem is highly nonlinear and generally nonconvex with respect to the model parameters \citep{Tarantola_1982_GNI,Symes_2008_MVA,Pratt_1998_GNF}. As a result, the data misfit function may possess multiple local minima, making the recovered solution strongly dependent on the initial model 
\citep[see, e.g.,][ and references therein for optimization methods tackling the local minima or cycle-skipping issue]{Operto_2023_FWI}.
In this paper, however, we assume that a sufficiently accurate background model $\bm_0$ is available and focus exclusively on estimating a model perturbation $\delta\bm$ such that the updated model, $\bm^*=\bm_0+\delta\bm$, provides an improved fit to the observed data.
Under this assumption, the nonlinear wave equation can be reformulated in terms of the background model and the model perturbation \citep{Abubakar_2009_FDC,Aghamiry_2021_OEF}.
In particular, the wave equation in \eqref{eq:forward_wave} can be decomposed into a linear term and a bilinear correction as \citep{Schuster_1985_HBB}
\begin{equation}  \label{Decomposition}
\mathbf{A}(\mathbf{m}_0)\mathbf{u}^*_s+
\left(\frac{\partial \mathbf{A(m)}}{\partial \mathbf{m}}\mathbf{u}^*_s
\right) \delta\mathbf{m}=\bb_s,
\end{equation}
where the first term describes wave propagation in the known background medium $\bm_0$, while the second term accounts for scattering induced by the model perturbation $\delta\mathbf{m}$.
Multiplying both sides of Eq.~\eqref{Decomposition} by $\mathbf{A}(\mathbf{m}_0)^{-1}$
and defining the bilinear mapping 
\begin{equation} \label{scs}
\varphi(\bu^*_s,\delta\bm) =
\left(-\frac{\partial \mathbf{A(m)}}{\partial \mathbf{m}}\bu^*_s\right) \delta\mathbf{m},
\end{equation}  
after rearranging the terms, the exact wavefield satisfies the Lippmann--Schwinger integral equation \citep{Prunty_2020_ALI} 
\begin{equation} \label{Lippmann}
\mathbf{u}^*_s =\bu^{\texttt{inc}}_s +\bA(\bm_0)^{-1}\varphi(\mathbf{u}^*_s,\delta\mathbf{m}),
\end{equation}
where $\bu^{\texttt{inc}}_s=\bA(\bm_0)^{-1}\mathbf{b}_s$ denotes the incident wavefield generated by the physical source $\mathbf{b}_s$ and propagating through the background model. 
$\varphi(\bu^*_s,\delta\bm)$ is referred to as the \emph{scattering source}, also known as the \emph{secondary Born source} \citep{Tarantola_1988_TBI} or the \emph{contrast source} \citep{vandenBerg_1997_CSI}. Physically, it represents secondary wavefields generated by the interaction between the wavefield and the model perturbation. Since both the wavefield and the perturbation are distributed throughout the computational domain, the scattering source generally occupies the entire space--time domain.
The mapping $\varphi(\mathbf{u},\delta\mathbf{m})$ is bilinear because it is linear with respect to the wavefield for a fixed perturbation, and linear with respect to the perturbation for a fixed wavefield, but nonlinear when both arguments are treated as unknowns simultaneously.
The Lippmann--Schwinger equation \eqref{Lippmann} is still nonlinear because the scattering source depends on the unknown total wavefield $\bu^{*}$. A common simplification is obtained by assuming that the model perturbation is sufficiently small so that multiple scattering can be neglected. Under this assumption, the total wavefield in the scattering source is replaced by the incident wavefield,
$\bu^{*}\approx\bu^{\texttt{inc}}$, which leads to the first-order Born approximation \citep{Schuster2017}.
Using this approximation in the scattering source,  
\eqref{Lippmann} yields the Born approximation of the wavefield,
\begin{equation}
\label{eq:born_wavefield}
\bu_s^{\texttt{Born}}
=\bu^{\texttt{inc}}_s + \bA(\bm_0)^{-1} \varphi(\bu_s^{\texttt{inc}},\delta\mathbf{m}).
\end{equation}
In this way, the bilinear mapping $\varphi$ becomes linear. Sampling the resulting wavefield at the receiver locations yields the linearized data equation
\begin{equation}
\label{eq:born_data}
\mathbf{d}^{\texttt{Born}}_s= \mathbf{d}_{0,s} + \bB_s\delta\bm, \quad s=1,...,N_s,
\end{equation}
where $\bd_{0,s}=\bP\bu^{\texttt{inc}}_s$ is the incident wavefield at the receivers, $\bB_s=\bP\bA(\bm_0)^{-1}\bL_s^{{\texttt{inc}}}$ is the Born modeling operator and $\bL_s^{{\texttt{inc}}} = -\frac{\partial \mathbf{A(m)}}{\partial \mathbf{m}}\bu^{\texttt{inc}}_s$ is the virtual source operator \citep{Pratt_1998_GNF}.  The operator $\bL_s^{{\texttt{inc}}}$ maps the model perturbation to the corresponding scattering source, $\varphi(\bu_s^{\texttt{inc}},\delta\bm)=\bL_s^{{\texttt{inc}}}\delta\bm$, while $\bA(\bm_0)^{-1}$ propagates this virtual source through the background medium and $\bP$ samples the resulting scattered wavefield at the receiver locations. Consequently, the Born modeling operator can be interpreted as the composition of three linear operators: scattering source generation, wave propagation, and receiver sampling.

%%%%%%%%%%%%%%%%%%%%%%%%%%%%%%%%%%%%%%%%%%%%%%%%%%%%%%
%%%%%%%%%%%%%%%%%%%%%%%%%%%% RTM + LSRTM
%%%%%%%%%%%%%%%%%%%%%%%%%%%%%%%%%%%%%%%%%%%%%%%%%%%%%%
\subsection{Reverse-Time Migration and Least-Squares Reverse-Time Migration}
Given the observed data, the objective is to estimate the model perturbation that best explains the data under the Born approximation.
This is done by solving \eqref{eq:born_data} in least squares form
\begin{equation}
\label{eq:LSRTM}
\minimize_{\delta\bm} ~J_{\texttt{LSRTM}}(\delta\bm),\quad\mbox{where}\quad J_{\texttt{LSRTM}}(\delta\bm):= \frac{1}{2}\sum_{s=1}^{N_s} \| \bB_s\delta\bm - \delta\mathbf{d}_s\|_2^2.
\end{equation}
Here, $\delta\mathbf{d}_s = \mathbf{d}_s-\mathbf{d}_{0,s}$ is the singly scattered data.
The LSRTM objective function $J_{\texttt{LSRTM}}$ can be interpreted as a Gauss--Newton quadratic approximation of the nonlinear FWI objective $J_{\texttt{FWI}}$ obtained by replacing the nonlinear forward operator with its first-order Taylor expansion around the background model \citep{Pratt_1998_GNF}. The Born modeling operator $\bB_s$ is precisely the Jacobian of the nonlinear forward modeling operator $G_s(\bm)$ evaluated at the background model,
$\bB_s =\left. \frac{\partial G_s(\bm)} {\partial\bm} \right|_{\bm_0}$. 
The gradient of the LSRTM objective function is therefore given by
\begin{equation}
\label{eq:LSRTM_grad}
\nabla J_{\texttt{LSRTM}}= \sum_{s=1}^{N_s}\bB_s^{\top}(\bB_s\delta\bm - \delta\mathbf{d}_s).
\end{equation}
Evaluating the negative gradient at the background model, $\delta\bm=\mathbf{0}$, yields the standard RTM image \citep{Baysal_1983_RTM, Whitmore_1983_IDM,Schuster2017},
\begin{equation}
\label{eq:RTM}
\delta\bm_{\texttt{RTM}}= \left. -\nabla J_{\texttt{LSRTM}}(\delta\bm) \right|_{\delta\bm=\mathbf{0}}
=\sum_{s=1}^{N_s}\bB_s^{\top}\delta\mathbf{d}_s
= \sum_{s=1}^{N_s} (\bL_s^{{\texttt{inc}}})^{\top} \bA(\bm_0)^{-\top} \bP^{\top} \delta\mathbf{d}_s.
\end{equation}
Here, the superscript $\top$ denotes matrix transposition (or, more generally, the adjoint operator). The operator $\bP^{\top}$ is the adjoint of the sampling operator and injects the scattered data at the receiver locations to form the adjoint sources. The operator $\bA(\bm_0)^{-\top}$ is the adjoint wave propagator, which back-propagates these adjoint sources through the background medium. Finally, the operator $({\bL_s^{{\texttt{inc}}}})^{\top}$ computes the zero-lag cross-correlation between the adjoint wavefield and the incident wavefield, producing the RTM image.

From the optimization point of view, since the Gauss--Newton quadratic model matches both the value and the gradient of the nonlinear FWI objective at the background model, the RTM image is identical to the negative gradient of the FWI objective evaluated at the same model. Consequently, RTM can be interpreted as the first search direction of classical gradient-based FWI \citep{Gauthier_1986_TDN}. In gradient-based FWI, the background model is iteratively updated, the forward operator is relinearized, and a new RTM search direction is computed until convergence. At each iteration, the model is updated by moving along the RTM search direction, scaled by a step length that is typically determined through a line-search procedure requiring additional forward wave-equation solves \citep{Pratt_1998_GNF,Schuster2017}.
Although RTM provides the steepest-descent direction of the quadratic objective, it ignores the coupling between model parameters introduced by the Born modeling operator. Consequently, the recovered image is affected by nonuniform illumination, unreliable amplitudes, acquisition geometry, and the limited bandwidth of the seismic source. LSRTM addresses these limitations by minimizing the quadratic objective function in \eqref{eq:LSRTM}, yielding the normal equations \citep{Nemeth_1999_LSQ, Dai2012,Zand_2020_COT,Davydenko2021FWLSRTM}
\begin{equation}
\label{eq:normal}
\left(
\sum_{s=1}^{N_s} \mathbf{B}_s^{\top}\mathbf{B}_s\right) \delta\bm_{\texttt{LSRTM}} = \delta\bm_{\texttt{RTM}}.
\end{equation}
The matrix
$\mathbf H_{\texttt{GN}} = \sum_{s=1}^{N_s} \mathbf{B}_s^{\top}\mathbf{B}_s$ 
is the Gauss--Newton approximation of the Hessian of the nonlinear FWI objective evaluated at the background model \citep{Pratt_1998_GNF}. Therefore, LSRTM can be interpreted as a Hessian-corrected version of RTM that compensates for nonuniform illumination and acquisition geometry by approximately inverting the Gauss--Newton Hessian.
If the resulting model perturbation $\delta\bm_{\texttt{LSRTM}}$ is used to update the background model and the Born modeling operator is subsequently relinearized, the repeated application of LSRTM recovers the classical Gauss--Newton algorithm for FWI \citep[e.g.,][]{Pratt_1998_GNF,Metivier_2017_TRU}.

In practice, the normal equations \eqref{eq:normal} are not solved by explicitly forming and inverting the Gauss--Newton Hessian, which is computationally prohibitive, but iteratively through a preconditioned gradient scheme, as summarized in Algorithm~\ref{ALG_LSRTM}. The full Gauss--Newton Hessian in \eqref{eq:normal} may be approximately inverted in several ways, ranging from wavenumber-domain and matrix-free iterative schemes to simple diagonal scalings \citep{Hu2001, Guitton2004, Metivier_2017_TRU}; here we adopt the diagonal pseudo-Hessian $\sum_{s=1}^{N_s} (\bL_s^{\texttt{inc}})^{\top}\bL_s^{\texttt{inc}}$, which is inexpensive to form from the incident wavefield and preconditions the gradient by compensating for the leading illumination and geometrical-spreading effects while discarding the off-diagonal coupling between model parameters. Because this preconditioner depends only on the incident wavefield, its diagonal $\mathbf{I}_{\texttt{DNUM}} = \sum_{s=1}^{N_s}\langle\partial_{tt}\bu_s^{\texttt{inc}},\partial_{tt}\bu_s^{\texttt{inc}}\rangle_t$ is accumulated once in the offline stage and reused throughout the iterations. Starting from a zero perturbation, each iteration back-propagates the current data residual $\delta\bd_s + \boldsymbol{\xi}_s^{k}$ to form the adjoint wavefield $\boldsymbol{\lambda}_s$ and correlates it with the incident wavefield to build the gradient $\mathbf{I}_{\texttt{NUM}} = \sum_{s=1}^{N_s}\langle\partial_{tt}\bu_s^{\texttt{inc}},\boldsymbol{\lambda}_s\rangle_t$; the model perturbation is then updated as $\delta\bm^{k+1} = \delta\bm^{k} - \gamma\,\mathbf{I}_{\texttt{NUM}}/\mathbf{I}_{\texttt{DNUM}}$ with a step length $\gamma$, which must be tuned in practice by one of several approaches, such as a line search or trial inversions, that can themselves add wave-equation solves per iteration. The updated perturbation is re-modeled through the Born operator to give the predicted scattered data $\boldsymbol{\xi}_s^{k+1} = -\bB_s\delta\bm^{k+1}$, which enters the residual that drives the next iteration; at the first iteration $\delta\bm^{0}=\mathbf{0}$ and $\boldsymbol{\xi}_s^{0}=\mathbf{0}$, so this residual reduces to the scattered data $\delta\bd_s$ and the update recovers the RTM image. Ideally, each iteration therefore entails one adjoint and one forward wave propagation per source, though in practice more may be required, as discussed in Section~\ref{sec:complexity}.
\begin{algorithm}[hbt!]
\caption{LSRTM in an acoustic medium}
\label{ALG_LSRTM}
\begin{algorithmic}[1]
\STATE \textbf{Input:} background model $\bm_0$, observed data $\{\bd_s\}$, sources $\{\bb_s\}$, step length $\gamma$
\STATE \textbf{Offline stage} (computed once):
\STATE Set $\mathbf{I}_{\texttt{DNUM}} = \mathbf{0}$
\FOR{$s=1,...,N_s$}
\STATE Compute $\bu_s^{\texttt{inc}} = \bA(\bm_0)^{-1}\bb_s$    \label{LSRTM_inc}
\STATE Compute $\delta\bd_s = \bd_s - \bP\bu_s^{\texttt{inc}}$
\STATE $\mathbf{I}_{\texttt{DNUM}} = \mathbf{I}_{\texttt{DNUM}} + \langle \partial_{tt}\bu_s^{\texttt{inc}},\partial_{tt}\bu_s^{\texttt{inc}}\rangle_t$
\ENDFOR
\STATE \textbf{Initialize:} Set $\delta\bm^{0} = \mathbf{0},~ \boldsymbol{\xi}_s^0=\mathbf{0}$, \; $k = 0$
\WHILE{convergence criterion not satisfied}
\STATE Set $\mathbf I_{\texttt{NUM}} = \mathbf{0}$
\FOR{$s=1,...,N_s$}
\STATE Compute $\boldsymbol{\lambda}_s$ by back-propagating $\delta\bd_s +\boldsymbol{\xi}_s^{k}$ in the background model $\bm_0$ \label{LSRTM_adj}
\STATE $\mathbf I_{\texttt{NUM}} = \mathbf I_{\texttt{NUM}} + \langle \partial_{tt}\bu_s^{\texttt{inc}},\boldsymbol{\lambda}_s\rangle_t$
\ENDFOR
\STATE Set $\delta\bm^{k+1} =\delta\bm^k -\gamma \frac{\mathbf I_{\texttt{NUM}}}{\mathbf I_{\texttt{DNUM}}}$
%%%%%%%%%%%%%%%%%%%%%%%%%%%%%%%
\FOR{$s=1,...,N_s$}
\STATE Compute the scattered data in the background model $\bm_0$: \label{LSRTM_for} \\
 $\boldsymbol{\xi}_s^{k+1} = - \bB_s\delta\bm^{k+1}$
\ENDFOR
\STATE $k \leftarrow k+1$
\ENDWHILE
\STATE \textbf{Output:} model perturbation $\delta\bm$
\end{algorithmic}
\end{algorithm}
\subsection{Challenges of LSRTM}

\subsubsection{Modeling challenges}

The first-order Born approximation replaces the total wavefield in the scattering source by the incident wavefield. Although this assumption linearizes the forward problem, it also neglects part of the scattering information contained in the exact wavefield. To characterize the missing information, let $\delta \bu = \bu^{*} - \bu^{\texttt{inc}}$ denote the total scattered wavefield, i.e., the difference between the exact and incident wavefields. Substituting $\bu^{*}=\bu^{\texttt{inc}}+\delta \bu$ into the Lippmann--Schwinger equation (\ref{Lippmann}) gives \citep{Gholami_2024_FWI}
\begin{equation} \label{Lippmann_2}
\mathbf{u}^*_s =
\underbrace{\bu^{\texttt{inc}}_s}_{\text{Incident}}
+\underbrace{\bA(\bm_0)^{-1}\varphi(\bu^{\texttt{inc}}_s,\delta\mathbf{m}) }_{\text{First-order Born scattering}}
+ \underbrace{\bA(\bm_0)^{-1}\varphi(\delta \bu_s,\delta\mathbf{m})}_{\text{Higher-order scattering}}.
\end{equation}
The first two terms on the right-hand side correspond to the incident and singly scattered wavefields, respectively, and together constitute the first-order Born approximation. The third term represents the contribution omitted by the Born approximation. Since it is generated by the interaction between the model perturbation and the scattered wavefield itself, it accounts for all higher-order scattering events, including multiple reflections and internal reverberations \citep{Operto_2023_FWI}.

Standard RTM and LSRTM reconstruct the image using only the incident and first-order scattered wavefields. While this approximation is accurate when the background model is sufficiently close to the true model, it may produce suboptimal images when higher-order scattering cannot be neglected.
The proposed DRTM, in contrast, seeks to recover the neglected higher-order scattering contribution. As will be shown in Section~\ref{sec:method}, this is achieved by formulating a fixed-point problem for the higher-order scattering term, which is iteratively reconstructed in the data space. This leads to an improved linearization of the nonlinear wave equation while preserving the computational efficiency and benefiting from the improved convexity of the linearized imaging objective functions.

\begin{figure}[htbp]
    \centering
    \includegraphics[width=0.6\linewidth]{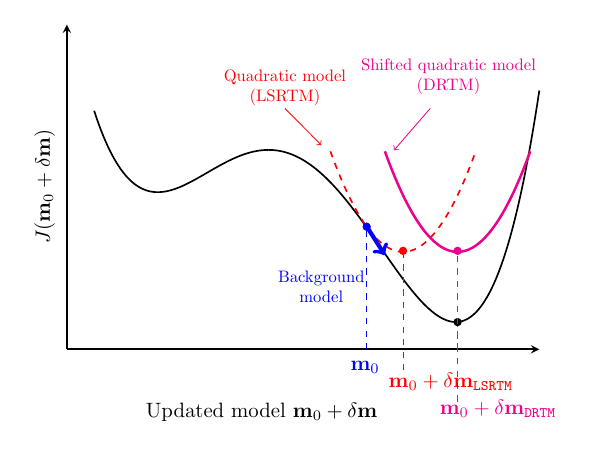}
    \caption{Schematic interpretation of DRTM. The black curve represents the data misfit objective function, while the dashed red curve denotes the Gauss--Newton quadratic model obtained by linearizing the forward operator about the background model $\bm_0$. Its minimizer corresponds to the updated model $\bm_0+\delta\bm_{\texttt{LSRTM}}$. By iteratively reconstructing the higher-order scattering contribution through the dual iteration, DRTM transforms this quadratic model into a shifted quadratic model (solid magenta), whose minimizer moves toward $\bm_0+\delta\bm_{\texttt{DRTM}}$. When the reconstructed correction exactly compensates for the neglected higher-order scattering, the minimizer of the multiplier-shifted quadratic model coincides with the global minimizer of the original nonlinear objective.
}
    \label{fig:obj_func}
\end{figure}

Figure~\ref{fig:obj_func} compares the nonlinear FWI objective with its quadratic approximation. The black curve represents the original nonconvex data-fitting objective, whereas the dashed red curve denotes the quadratic model obtained from the first-order Born approximation. Owing to the neglected higher-order scattering, the minimizer of this quadratic model generally differs from that of the nonlinear objective. The proposed DRTM modifies this quadratic model by iteratively reconstructing the missing higher-order scattering contribution, leading to the shifted quadratic model illustrated by the solid magenta curve.

\subsubsection{Computational and Algorithmic challenges}
In addition to the modeling limitations discussed above, LSRTM poses computational and algorithmic challenges when applied to realistic seismic imaging problems. In principle, the adjoint operator could be constructed explicitly by first assembling the Born modeling matrix $\bB_s$ and then forming its transpose. Such a matrix-based approach provides a straightforward means of obtaining the exact adjoint operator and is feasible for small-scale problems \citep{Yao_2016_LSRTM}. However, for realistic seismic imaging problems, where the number of model parameters may reach millions or billions, the explicit construction and storage of $\bB_s$ become computationally prohibitive. Consequently, both the forward and adjoint Born operators must be implemented in a matrix-free manner by exploiting the underlying wave-equation operators \citep{Metivier_2017_TRU,Chen_2020_TGN}. In a matrix-free implementation, the forward product $\bB_s\bp$ is computed by generating a virtual source, propagating the resulting scattered wavefield through the background medium, and sampling it at the receiver locations. Conversely, the adjoint product $\bB_s^\top \bq$ is obtained by injecting the data residual at the receivers, reverse-time propagating the adjoint wavefield, and correlating it with the incident wavefield.
Consequently, every iteration of an iterative LSRTM solver requires, for each source, multiple forward and adjoint wave-equation solves, which together constitute the dominant computational cost.
{Another practical challenge is the implementation of the adjoint operator. Since matrix-free optimization relies on the mathematical identity $\langle \mathbf{B}_s\bp,\bq\rangle=\langle \bp,\mathbf{B}_s^{\top}\bq\rangle$,
the adjoint operator must be implemented consistently with the Born forward modeling operator. Adjoint tests are therefore routinely employed to verify the correctness of large-scale LSRTM implementations.}

These modeling and computational challenges motivate the development of the proposed imaging method. From a modeling perspective, DRTM iteratively reconstructs the higher-order scattering information neglected by the first-order Born approximation, leading to a progressively improved linearization of the nonlinear wave equation. From a computational perspective, DRTM adopts a hybrid time--frequency formulation in which the Gauss--Newton Hessian is represented in the data space rather than the model space. Since the data-space Hessian is assembled independently for each frequency and is typically several orders of magnitude smaller than its model-space counterpart, it can be formed explicitly and efficiently inverted using standard dense linear algebra techniques. Consequently, DRTM eliminates the need for large-scale matrix-free Hessian solvers while reducing the computational cost of multi-source seismic imaging.

\section{Dual Reverse Time Migration} \label{sec:method}

The previous section showed that the principal limitation of LSRTM stems from replacing the true scattering source with its first-order Born approximation. Consequently, improving the Born linearization amounts to recovering the higher-order scattering contribution omitted by the conventional LSRTM formulation. Rather than repeatedly relinearizing the forward operator, as in Gauss--Newton FWI, we seek to iteratively reconstruct this missing scattering information while preserving the background model.
To achieve this objective, we revisit the Lippmann--Schwinger equation \eqref{Lippmann_2} and reformulate the problem in terms of the missing scattering source. Recall that the exact scattered wavefield is generated by the scattering source
$\varphi(\mathbf{u}_s^{*},\delta\mathbf{m})$, 
which contains both the first-order Born term and all higher-order scattering contributions. We therefore introduce the dual scattering source
\begin{equation}\label{qe:constitutive}
    \blambda_s = \varphi(\bu_s,\delta\mathbf{m}),
\end{equation}
as an independent unknown.
Suppose that $\bu_s$ and $\boldsymbol{\lambda}_s$ denote current estimates of the wavefield and the corresponding dual scattering source, respectively. They must satisfy the constitutive equation \eqref{qe:constitutive}. Measuring its violation, we introduce the dual residual
\begin{equation}\label{epsilon}
    \boldsymbol{\varepsilon}_s=\blambda_s-\varphi(\bu_s,\delta\mathbf{m}).
\end{equation}
The Lippmann--Schwinger equation (\ref{Lippmann_2}) can now be written as 
\begin{equation} \label{eq:state}
\mathbf{u}_s =\bu^{\texttt{inc}}_s
+\bA(\bm_0)^{-1}(\blambda_s-\boldsymbol{\varepsilon}_s).
\end{equation}
Sampling the state equation \eqref{eq:state} at the receiver locations gives
\begin{equation} 
\bP\mathbf{u}_s =\bP\bu^{\texttt{inc}}_s +\bP\bA(\bm_0)^{-1}(\blambda_s-\boldsymbol{\varepsilon}_s).
\end{equation}
Introducing the scattered data $\delta \bd_s = \bd_s-\bP\bu^{\texttt{inc}}_s$ and defining $\bS=\bP\bA(\bm_0)^{-1}$, the data equation becomes
\begin{equation} \label{eq:data_dual}
\delta \bd_s =\bS[\blambda_s-\boldsymbol{\varepsilon}_s].
\end{equation}

Equations \eqref{qe:constitutive}, \eqref{eq:state}, and \eqref{eq:data_dual} form the foundation of the proposed DRTM formulation. The constitutive relation links the dual scattering source to the wavefield and the model perturbation, the state equation governs wave propagation in the background medium, and the data equation enforces consistency with the observed scattered data.
Starting from an initial estimate of the dual residual, for example $\boldsymbol{\varepsilon}_s=\mathbf{0}$, the data equation is first solved to estimate the dual scattering source. The recovered dual scattering source is then used in the state equation to reconstruct the wavefield. Together, the reconstructed wavefield and dual scattering source provide an updated estimate of the model perturbation through the constitutive relation. Finally, the dual residual is recomputed from \eqref{epsilon}, and the process is repeated until convergence. In this manner, the higher-order scattering information omitted by the first-order Born approximation is progressively reconstructed while keeping the background model fixed \citep{Gholami_2024_FWI,Aghazade_2025_FAF}.

The main challenge in applying this method in the time domain is that the dual variables $\boldsymbol{\varepsilon}_s$ reside in the wavefield space. For realistic seismic imaging problems, explicitly storing and updating these variables for all source experiments requires a prohibitive amount of memory. To overcome this limitation, we project the dual variables into the data space and reformulate the fixed-point iteration in terms of the corresponding higher-order scattering data.
We start by solving \eqref{eq:data_dual} for $\boldsymbol{\lambda}_s$ using the following damped least-squares problem
\begin{equation}
    \boldsymbol{\lambda}_s = \arg\min_{\boldsymbol{\lambda}}
    \|\bS\boldsymbol{\lambda} - (\delta\mathbf{d}_s+\bS\boldsymbol{\varepsilon}_s)\|_2^2 + \mu\|\boldsymbol{\lambda}\|_2^2, \quad s=1,...,N_s.
\end{equation}
This is a quadratic optimization problem that admits the closed-form solution 
\begin{equation}
\label{eq:lambda_closed}
\boldsymbol{\lambda}_s
= \bS^{\top} \bQ^{-1} \left(\delta\mathbf{d}_s+\bS\boldsymbol{\varepsilon}_s\right), \quad s=1,...,N_s,
\end{equation}
where $\bQ=\bS\bS^{\top}+\mu\bI$ is the regularized data-space Hessian and $\mu>0$ is the penalty parameter.
We project $\boldsymbol{\varepsilon}_s$ into the data space by introducing the data-space dual variable $\boldsymbol{\nu}_s=\bQ^{-1}\bS\boldsymbol{\varepsilon}_s$. Then with the approximation $\bS^{\top} \bQ^{-1}\bS \approx \bI$, the state equation becomes
\begin{equation} \label{eq:state2}
\mathbf{u}_s =\bu^{\texttt{inc}}_s
+\bA(\bm_0)^{-1} \bS^{\top}  \bQ^{-1}\delta\mathbf{d}_s.
\end{equation}

The data-space dual variable is updated \citep{Gholami_2022_EFW} as
\begin{equation} \label{nu_update}
\boldsymbol{\nu}^{k+1}_s \leftarrow \boldsymbol{\nu}^{k}_s + \bQ^{-1}\left(\bd_s - \bP \bA(\bm_0 + \delta \bm(\boldsymbol{\nu}^{k}))^{-1} \bb_s\right),
\end{equation}
where $\delta \bm(\boldsymbol{\nu}^{k})$ denotes the associated model perturbation computed online using $\mathbf{u}_s$ and $\boldsymbol{\lambda}_s$. Algorithm~\ref{ALG_DRTM} summarizes the implementation of DRTM in detail.
Since the right-hand side of Eq.~\eqref{nu_update} depends only on the variable $\boldsymbol{\nu}^{k}$, the update defines a fixed-point iteration,
\begin{equation} \label{eq:FP}
    \boldsymbol{\nu}^{k+1} = g(\boldsymbol{\nu}^k),
\end{equation}
where
$\boldsymbol{\nu}= 
    \begin{bmatrix}
        \boldsymbol{\nu}_1& \boldsymbol{\nu}_2& \cdots& \boldsymbol{\nu}_{N_s}
    \end{bmatrix}$
and $g$ denotes the corresponding fixed-point mapping. The evaluation of $g$ separates into an offline stage, which depends only on the fixed background model and is computed once before the iterations begin, and an online stage, which is repeated at every iteration.

\begin{figure}[htbp]
    \centering
\includegraphics[width=1.0\linewidth]{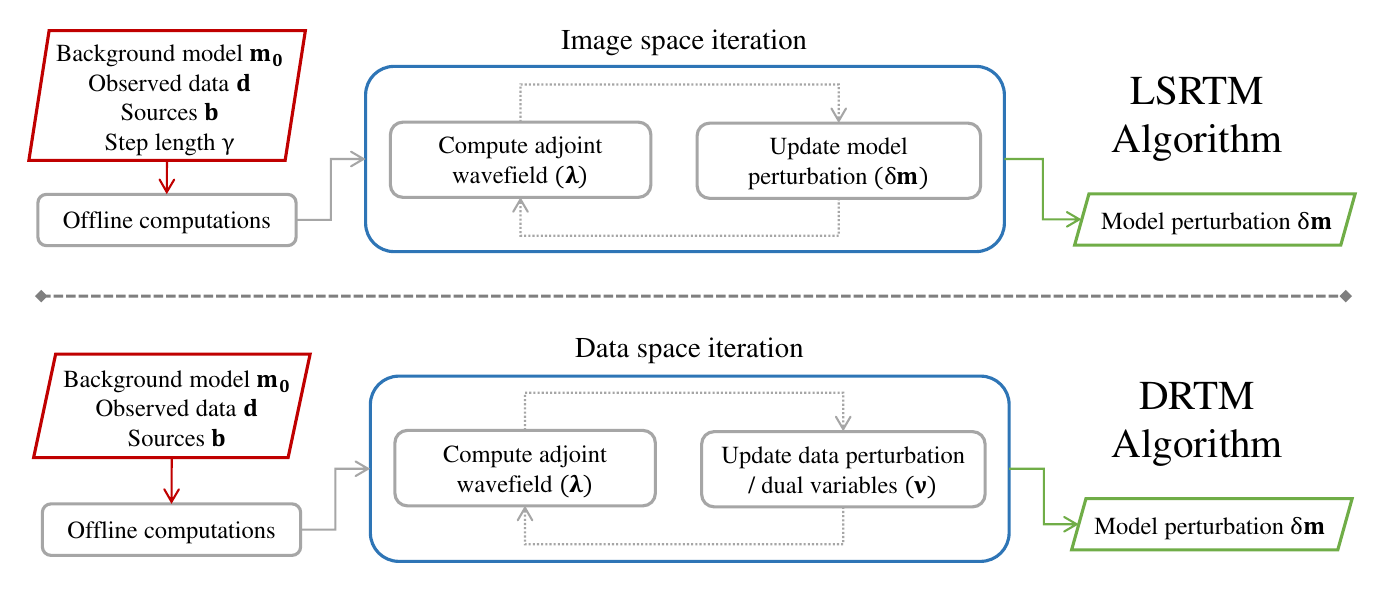}
    \caption{Comparison of the main steps of the LSRTM and DRTM methods.}
    \label{fig:flowchart}
\end{figure}

Figure~\ref{fig:flowchart} shows a schematic comparison of the main steps of the LSRTM and DRTM methods.

%%%%%%%%%%%%%%%%%%%%%%%%%%%%%%%%%%%%%%%%%%%%%%%%%%%%%%%%%%%%%%%%%%%%%%%%%%%%%%%%%%%%%
\begin{algorithm}[hbt!]
\caption{Dual reverse time migration in an acoustic medium}
\label{ALG_DRTM}
\begin{algorithmic}[1]
\STATE \textbf{Input:} background model $\bm_0$, observed data $\{\bd_s\}$, sources $\{\bb_s\}$
\STATE \textbf{Offline stage} (independent of $\boldsymbol{\nu}$, computed once):
\STATE Compute data-space Hessian $\bQ$ (Appendix~\ref{app:hessian})
\STATE Set $\mathbf I_{\texttt{DNUM}}=\mathbf{0}$ 
\FOR{$s=1,...,N_s$}
\STATE Compute $\bu_s^{\texttt{inc}} = \bA(\bm_0)^{-1}\bb_s$  \label{DRTM_wf_start}
\STATE Compute $\delta\bd_s = \bd_s - \bP\bu_s^{\texttt{inc}}$
\STATE Compute $\delta\bd_s^{e} = \bQ^{-1}\delta\bd_s$
\STATE Compute $\boldsymbol{\lambda}_s$ by back-propagating $\delta\bd_s^{e}$ in the background model $\bm_0$
\STATE Compute $\delta \bu_s=\bA(\bm_0)^{-1}\boldsymbol{\lambda}_s$ and set $\bu_s = \bu_s^{\texttt{inc}} + \delta \bu_s$  \label{DRTM_wf_end}
\STATE $\mathbf I_{\texttt{DNUM}} = \mathbf I_{\texttt{DNUM}} + \langle \partial_{tt}\bu_s,\partial_{tt}\bu_s\rangle_t$
\ENDFOR
\STATE \textbf{Initialize:} Set $\boldsymbol{\nu}_s^{0} = \mathbf{0}$, \; $k = 0$
\WHILE{convergence criterion not satisfied}
\STATE Set $\mathbf I_{\texttt{NUM}} = \mathbf{0}$
\FOR{$s=1,...,N_s$}
\STATE Compute $\boldsymbol{\lambda}_s$ by back-propagating $\delta\bd_s^{e}+\boldsymbol{\nu}_s^{k}$ in the background model $\bm_0$ \label{DRTM_adj}
\STATE $\mathbf I_{\texttt{NUM}} = \mathbf I_{\texttt{NUM}} + \langle \partial_{tt}\bu_s,\boldsymbol{\lambda}_s\rangle_t$
\ENDFOR
\STATE Set $\delta\bm^{k} = -\, \frac{\mathbf I_{\texttt{NUM}}}{\mathbf I_{\texttt{DNUM}}}$ \label{DRTM_dm}
%%%%%%%%%%%%%%%%%%%%%%%%%%%%%%%
\FOR{$s=1,...,N_s$}
\STATE Compute the data residual in the perturbed model $\bm_0+\delta\bm^{k}$: \\
$\delta\bd_s = \bd_s - \bP\bA(\bm_0+\delta\bm^{k})^{-1}\bb_s$  \label{DRTM_for}
\STATE     
$\boldsymbol{\nu}_s^{k+1}= \boldsymbol{\nu}_s^{k} + \bQ^{-1} \delta\bd_s$  \label{DRTM_dual}
\ENDFOR
\STATE $k \leftarrow k+1$
\ENDWHILE
\STATE \textbf{Output:} Model perturbation $\delta\bm$
\end{algorithmic}
\end{algorithm}

%%%%%%%%%%%%%%%%%%%%%%%%%%%%%%%%%%%%%%%%%%%%%%%%%%%%%%%%%%%%%%%%%%%%%%%%%%%%%%%%%%%%%
\section{Computational cost} \label{sec:complexity}
In this paper, it is assumed that the forward-propagating wavefield is computed once and stored at the Nyquist rate; otherwise, other suitable methods, such as checkpointing and recomputation, are necessary \citep{Symes_2007_RTM}. Under this assumption, the implementation of RTM requires $2N_s$ PDE solves ($N_s$ forward plus $N_s$ adjoint solves).
LSRTM requires additional PDE solves due to its iterative nature. {Before the iterations, the LSRTM method requires the incident wavefields to be calculated (Algorithm~\ref{ALG_LSRTM}, Line~\ref{LSRTM_inc}), resulting in an offline cost of $N_s$ PDE solves.} At each iteration, one forward (Algorithm~\ref{ALG_LSRTM}, Line~\ref{LSRTM_for}) and one backward (Algorithm~\ref{ALG_LSRTM}, Line~\ref{LSRTM_adj}) wavefield computation are required. Therefore, assuming $N_{\texttt{iter}}$ iterations, these two propagations amount to $2N_s N_{\texttt{iter}}$ PDE solves.
Furthermore, the Born modeling operator essentially propagates $\bL_s^{{\texttt{inc}}}\delta\bm$. Although this quantity can be computed from the stored wavefields using methods such as interpolation, in this paper we regenerate it on the fly, requiring one additional PDE solve per source per iteration. This results in a total online cost of $3N_sN_{\texttt{iter}}$ PDE solves.
An additional cost is also incurred for the computation of the step length $\gamma$ \citep{Pica_1990_NIS, Liu_2016_ESL}.

For the DRTM method, before the iterations, the data-space Hessian $\bQ$ and the forward wavefield need to be computed. Computing the data-space Hessian $\bQ$ requires $N_r$ PDE solves (Appendix~\ref{app:hessian}) and the forward wavefield requires $3N_s$ PDE solves (Algorithm~\ref{ALG_DRTM}, Lines~\ref{DRTM_wf_start}--\ref{DRTM_wf_end}), resulting in a total offline cost of $N_r + 3N_s$ PDE solves.
At each iteration of the DRTM method, one forward (Algorithm~\ref{ALG_DRTM}, Line~\ref{DRTM_for}) and one backward (Algorithm~\ref{ALG_DRTM}, Line~\ref{DRTM_adj}) wavefield computation are required. 
It should be noted that the DRTM method computes the data by solving the full wave equation in the perturbed model; thus, only one forward PDE solve is required, unlike Born modeling, which requires two PDE solves to compute the data. Therefore, the total online cost is $2N_s N_{\texttt{iter}}$ PDE solves.
Also, it is worth noting that in this algorithm, no additional cost is needed for step-length tuning.

The computational costs of the different methods are compared in Table~\ref{tab:cost}, and the actual number of PDE solves performed by LSRTM and DRTM methods for each model in Section~\ref{sec:results} is listed in Table~\ref{tab:pde_counts}.
 
\begin{table}[htbp]
\centering
\caption{Computational cost of RTM, LSRTM, and DRTM.}
\label{tab:cost}
\begin{tabular}{lcccc}
\hline
Method & Offline & Online & Additional requirement & Total \\
\hline
RTM & $2N_s$ & --- & --- & $2N_s$ \\
LSRTM & $N_s$ & $3N_sN_{\texttt{iter}}$ & step-length tuning & $N_s + 3N_sN_{\texttt{iter}}$ \\
DRTM & $N_r + 3N_s$ & $2N_sN_{\texttt{iter}}$ & --- & $N_r + 3N_s + 2N_sN_{\texttt{iter}}$ \\
\hline
\end{tabular}
\end{table}

\begin{table}[htbp]
\centering
\caption{Actual number of PDE solves performed for the models in Section~\ref{sec:results}.}
\label{tab:pde_counts}
\begin{tabular}{|c|c|c|c|c|c|}
\hline
Model & \begin{tabular}[c]{@{}c@{}}Reciprocity\\ receivers\end{tabular} & \begin{tabular}[c]{@{}c@{}}Reciprocity\\ sources\end{tabular} & Iterations & \begin{tabular}[c]{@{}c@{}}LSRTM PDEs\\ (Offline + Online)\end{tabular} & \begin{tabular}[c]{@{}c@{}}DRTM PDEs\\ (Offline + Online)\end{tabular} \\ \hline
\begin{tabular}[c]{@{}c@{}}Marmousi\end{tabular} & 192 & 31 & 50 & $31 + 4650 = 4681$ & $285 + 3100 = 3385$ \\ \hline
GO\_3D\_OBS & 778 & 78 & 50 & --- & $1012 + 7800 = 8812$ \\ \hline
\end{tabular}
\end{table}

%%%%%%%%%%%%%%%%%%%%%%%%%%%%%%%%%%%%%%%%%%%%%%%%%%%%%%%%%%%%%%%%%%%%%%%%%%%%%%%%%%%%%

\section{Numerical Results} \label{sec:results}
We evaluate the DRTM method on two benchmark models: the Marmousi model \citep{Versteeg_1994_Marmousi} and a crustal-scale model \citep{Gorszczyk_2021_MBG}. 
In all experiments, the water-layer parameters are assumed to be known; the model parameters were constrained to lie between the minimum and maximum values of the true reflectivity models, and no model regularization methods were used. The parameter $\mu$ in the data-space Hessian was set to a fraction, between $10^{-5}$ and $10^{-3}$, of the maximum eigenvalue of $\bS\bS^{\top}$, for each experiment. Convergence of DRTM and LSRTM methods was accelerated by using the Anderson acceleration method~\citep{Walker_2011_AAF, Aghazade_2022_AAA}.

%%%%%%%%%%%%%%%%%%%%%%%%%%%%%%%%%%%%%%%%%%%%%%%%%%%%%%%%%%%%%%%%%%%%%%%%%%%%%%%%%%%%%
\subsection{Marmousi model}
\begin{figure}[htbp]
    \centering
    \includegraphics[width=0.7\linewidth]{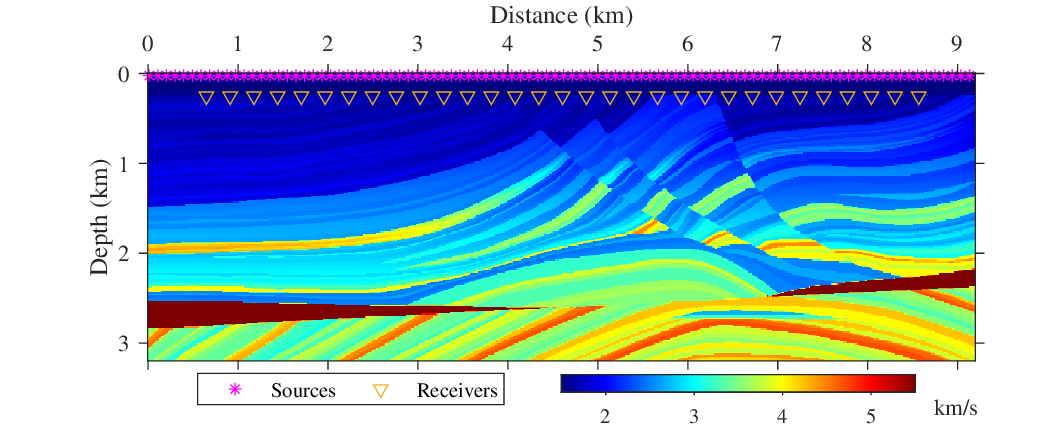}
    \caption{Marmousi true velocity model with the acquisition geometry overlaid: receivers are shown as orange triangles and sources as purple stars.}
    \label{fig:marm_m}
\end{figure}

\begin{figure}[htbp]
    \centering
    \includegraphics[width=0.9\linewidth]{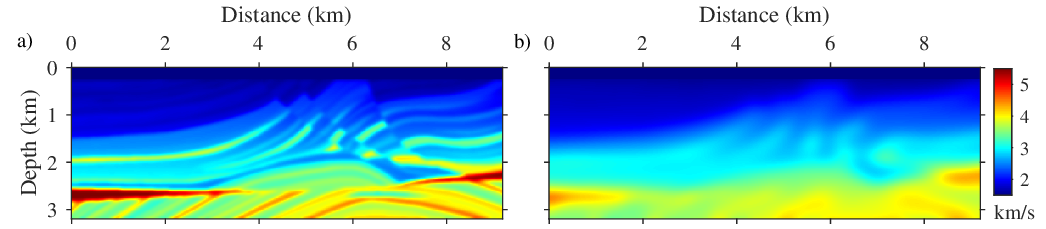}
    \caption{Migration velocity models generated by smoothing the true velocity model in Figure~\ref{fig:marm_m} using two different windows: (a) background model 1 obtained with a Gaussian window of 60~m and (b) background model 2 obtained with a Gaussian window of 240~m.}
    \label{fig:marm_m0}
\end{figure}

The Marmousi model spans approximately 9~km in distance and 3~km in depth (Figure~\ref{fig:marm_m}). It was discretized with a spatial sampling interval of 12~m. The acquisition consisted of 192 airgun sources at the surface, spaced 48~m apart, and 31 OBSs spaced 264~m apart, with a recording time window of 4.5~s and a sampling interval of 1~ms. The receivers extend from 648~m to 8568~m in distance, leaving approximately 636~m at each edge of the model without receiver coverage. The source wavelet was a 10~Hz Ricker wavelet. Using reciprocity, sources were treated as receivers and vice versa. The uncovered edges provide a natural test of each method's ability to maintain reflector continuity beneath poorly illuminated zones. Two migration  (background) velocity models were generated by smoothing the true velocity model with Gaussian windows of 60~m (Figure~\ref{fig:marm_m0}(a)) and 240~m (Figure~\ref{fig:marm_m0}(b)); we refer to them as background models 1 and 2, respectively.

\begin{figure}[htbp]
    \centering
    \includegraphics[width=0.8\linewidth]{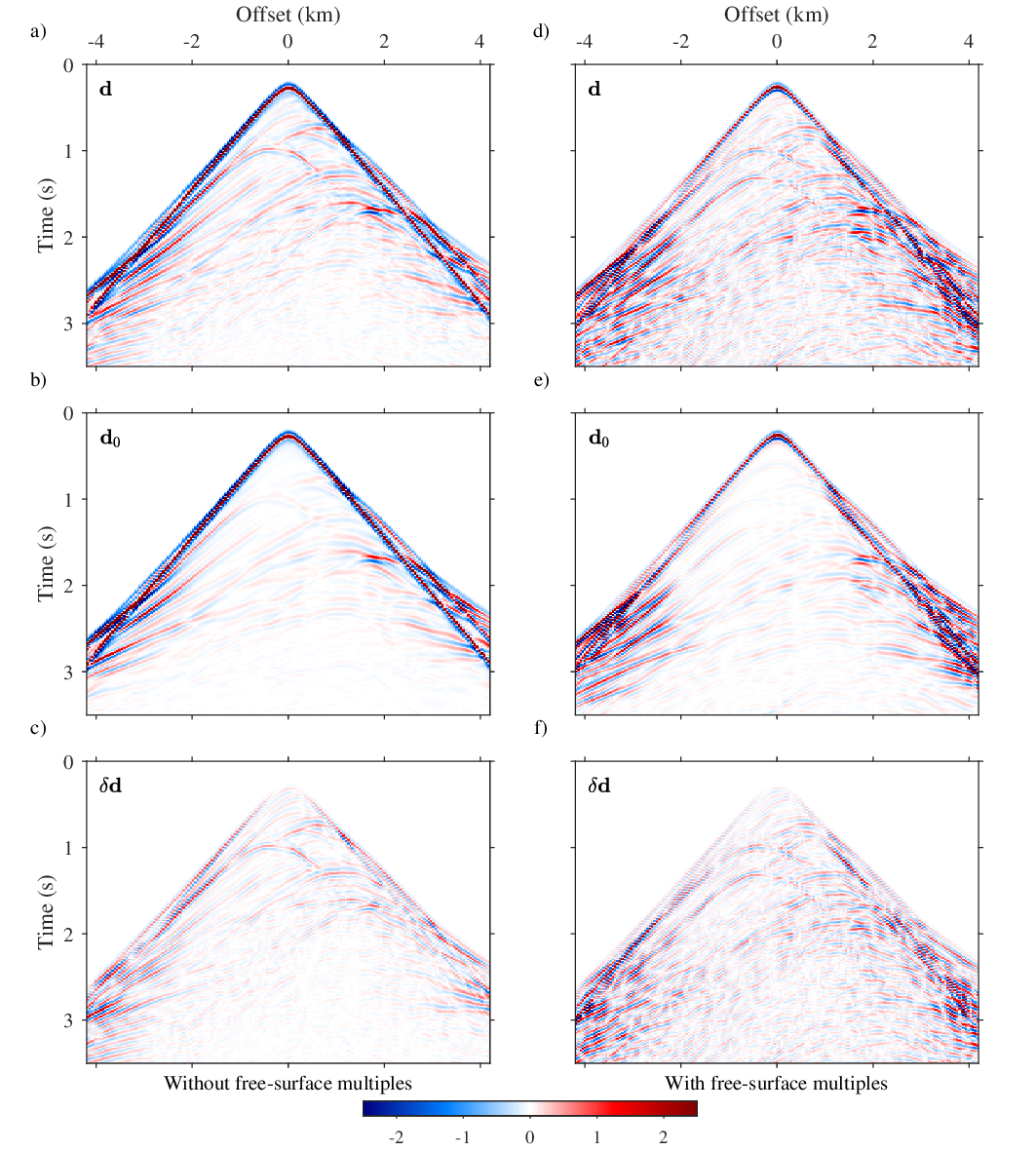}
    \caption{Representative common-shot gathers from the Marmousi model. (a--b) Data simulated in the (a) true model and (b) background model 1 (Figure~\ref{fig:marm_m0}(a)), without free-surface multiples. (c) Data residual (the difference between (a) and (b)). (d--f) The same as (a--c), respectively, but simulated with free-surface multiples.}
    \label{fig:marm_data}
\end{figure}

The data were simulated in two ways: without and with free-surface multiples. In the former case, absorbing boundary conditions were applied along all model boundaries, preventing the generation of free-surface multiples. In practice, such multiples are treated as noise and excluded from the data by assuming that they have been removed during preprocessing using methods such as surface-related multiple elimination (SRME) \citep{verschuur1992adaptive}. Figure~\ref{fig:marm_data}(a) shows a representative common-shot gather from the Marmousi model without free-surface multiples, while Figure~\ref{fig:marm_data}(b) shows the corresponding data simulated using background model 1 (Figure~\ref{fig:marm_m0}(a)). Figure~\ref{fig:marm_data}(c) shows the data residual as the input for imaging. 
For the free-surface simulations, the top boundary was treated as a free surface, while absorbing boundary conditions were applied along the other three sides of the model. Figures~\ref{fig:marm_data}(d)--(f) show the common-shot gathers from the true model, background model 1, and their difference, respectively, with free-surface multiples. In this case, the resulting free-surface multiples were retained and used as part of the imaging signal.

\begin{figure}[htbp]
    \centering
    \includegraphics[width=0.45\linewidth]{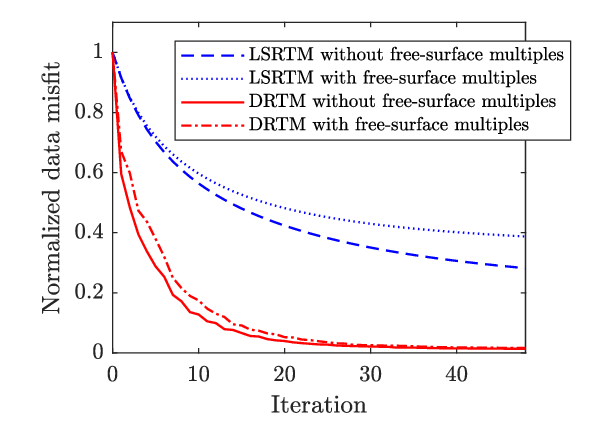}
    \caption{Normalized data misfit versus iteration for LSRTM and DRTM using background model 1, without and with free-surface multiples.}
    \label{fig:marm_curve60}
\end{figure}
%%--------------------------------------------------------------------------
\begin{figure}[htbp]
    \centering
    \includegraphics[width=0.9\linewidth]{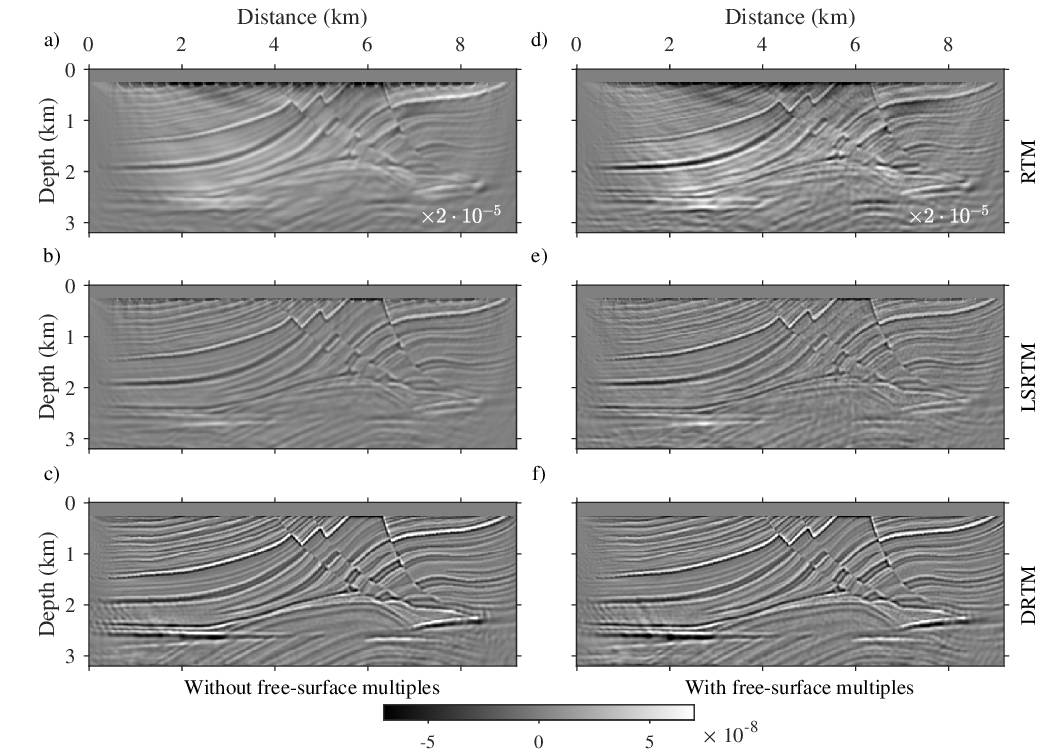}
    \caption{(a--c) Images obtained from the Marmousi data using (a) RTM, (b) LSRTM, and (c) DRTM, without free-surface multiples, using background model 1 shown in Figure~\ref{fig:marm_m0}(a). (d--f) Same as (a--c), but with free-surface multiples. All panels share a common color scale.}
    \label{fig:marm_results}
\end{figure}

We applied the RTM, LSRTM, and DRTM methods under an identical setup for both scenarios. The LSRTM and DRTM results were obtained after 50 iterations. Anderson acceleration with a memory of 6 was employed to improve convergence. 
Figure~\ref{fig:marm_curve60} shows the associated convergence curves.
Figure~\ref{fig:marm_results} compares the resulting images. Figures~\ref{fig:marm_results}(a)--(c) show the results corresponding to the case without surface-related multiples, whereas Figures~\ref{fig:marm_results}(d)--(f) show the corresponding results obtained using data with free-surface multiples.
All image panels share a unit color scale, fixed to the amplitude of the true model perturbation, so that they can be compared directly in both structure and amplitude. 

We see from Figures~\ref{fig:marm_results}(a) and \ref{fig:marm_results}(d) that RTM reconstructs the strong reflectors, but suffers from a pronounced acquisition footprint in the shallow section, rapid amplitude decay with depth, and poor focusing of the steeply dipping fault system between roughly 5 and 7~km. Moreover, in the deeper areas, the anticline beneath the fault system and the lateral salt creep, along with the associated geological events and the hydrocarbon reservoir beneath it, are poorly reconstructed. In Figures~\ref{fig:marm_results}(b) and \ref{fig:marm_results}(e), LSRTM suppresses the acquisition footprint and the low-frequency noise and improves the resolution, but its improvement in the deeper, structurally complex zones remains weak, such as within the anticline and beneath the salt creep. This is consistent with the role of LSRTM as a migration that differs from RTM only in a more accurate inversion of the model-space Hessian, which corrects illumination and amplitude but cannot introduce energy beyond the single-scattering regime.
The DRTM results (Figures~\ref{fig:marm_results}(c) and \ref{fig:marm_results}(f)) show improvements in these challenging regions. Relative to LSRTM, the resolution is enhanced throughout the model. The deep structures, the anticline, and the salt creep, as well as the reservoir and reflectors beneath the salt creep, are reconstructed. In addition, the fault structure at 5--7~km distance is defined with enhanced continuity.
The improvement is especially pronounced at both uncovered edges of the model, where the truncated receiver array leaves the shallow-to-deep section illuminated only by energy propagating in through multiple scattering. RTM and LSRTM both lose reflector continuity in these regions because, as single-scattering schemes, they are unable to image multiples, whereas the DRTM method recovers more continuous and interpretable reflectors there by leveraging the imaging of internal multiples, despite the identical acquisition.
Another notable improvement of DRTM over RTM and LSRTM is the successful elimination of the acquisition footprint and the reconstruction of high-resolution shallow reflectors. DRTM achieves this through the multi-scattering nature of its dual operator, which enables imaging of internal multiples by propagating the wavefield through the full wave equation. 

Furthermore, a comparison of Figures~\ref{fig:marm_results}(a)--(c) and (d)--(f) shows that incorporating free-surface multiples improves the image quality for all three methods. This improvement is more pronounced on both sides of the model, where illumination is weak due to the limited acquisition geometry.
For DRTM, the improvement is less pronounced than for the other two methods. This can be explained by the relatively accurate background model used in this example; the DRTM result is already of good quality due to its use of internal multiples. In contrast, the RTM and LSRTM images do not benefit from internal multiples. Incorporating surface multiples therefore improves their images by reducing footprint noise and enhancing shallow reflectors. Moreover, because these approaches are based on single scattering, only the source-ghost multiples among the different orders of surface-related multiples can contribute to the images.
The main drawback of incorporating surface-related multiples in these methods is the emergence of spurious near-vertical events, most clearly around 5 and 6~km distance, extending from approximately 3~km depth to about 1~km. These events do not correspond to real reflectors but are likely caused by misplaced multiple energy or crosstalk noise.

\begin{figure}[htbp]
    \centering
    \includegraphics[width=1\linewidth]{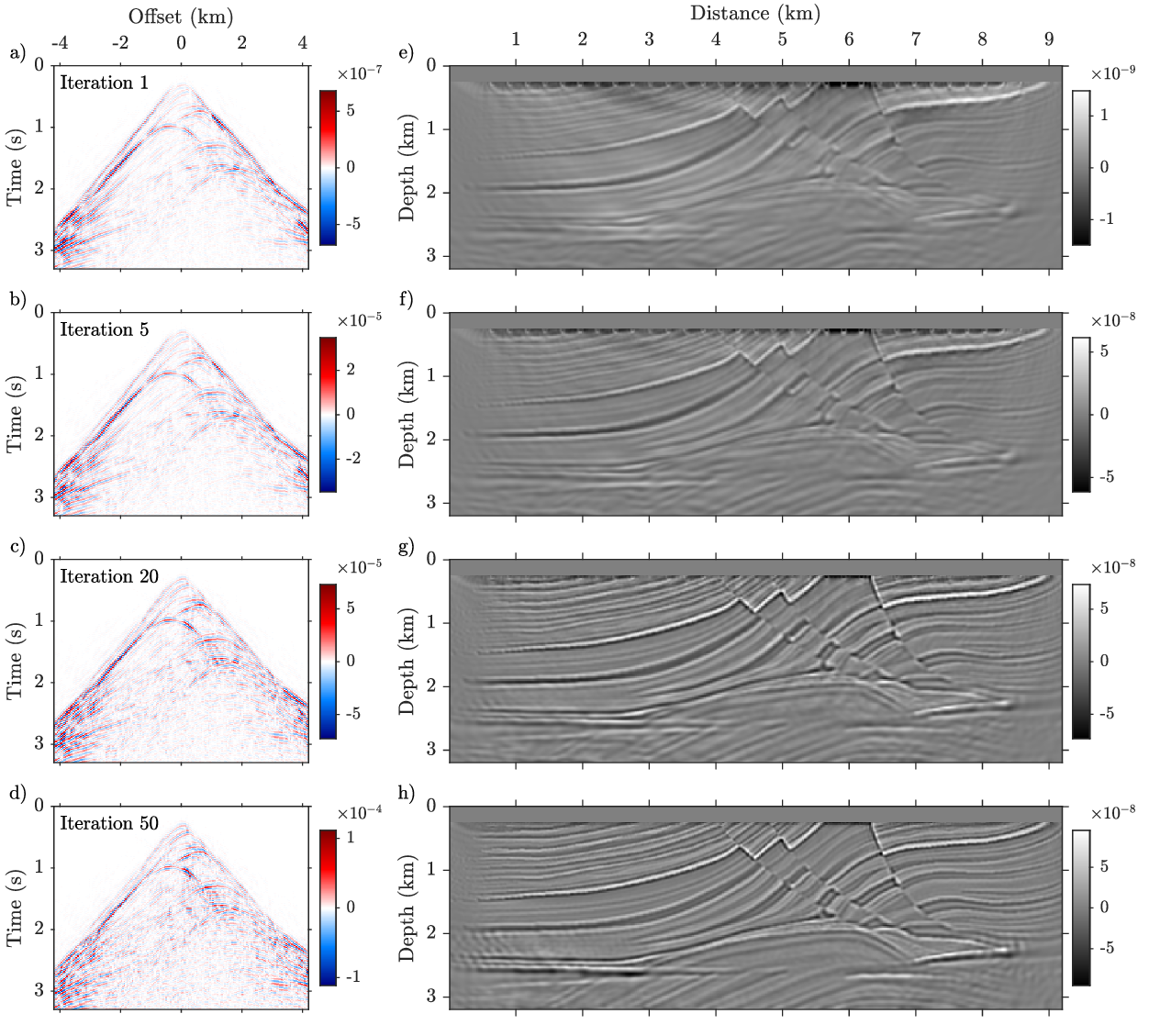}
    \caption{(a--d) Dual data for the case of no free-surface multiples at iterations (a) 1, (b) 5, (c) 20, and (d) 50. (e--h) The migration images obtained from the dual data in (a--d), respectively.}
    \label{fig:marm_duals_WoFS}
\end{figure}

\begin{figure}[htbp]
    \centering
    \includegraphics[width=1\linewidth]{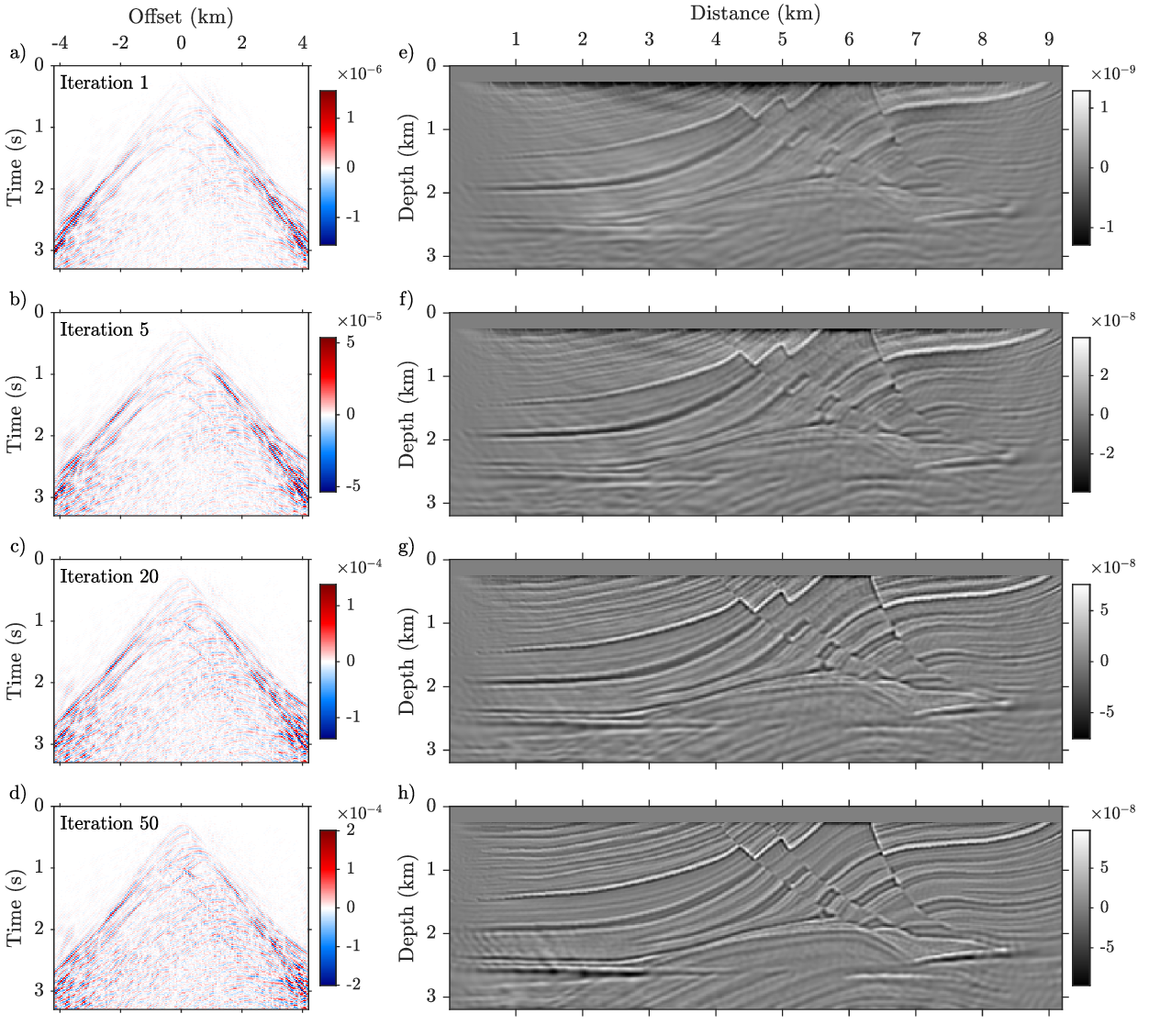}
    \caption{(a--d) Dual data for the case with free-surface multiples at iterations (a) 1, (b) 5, (c) 20, and (d) 50. (e--h) The migration images obtained from the dual data in (a--d), respectively.}
    \label{fig:marm_duals_WFS}
\end{figure}

Figures~\ref{fig:marm_duals_WoFS}(a)--(d) show the dual data estimated by the DRTM algorithm at different iterations (1, 5, 20, 50). These dual data contain key information---including all orders of multiple scattering and the effect of the inverse Hessian---required for high-resolution imaging. As the iterations progress, increasingly more information is incorporated into the estimated dual data. This progressive incorporation of information is more clearly visible in the corresponding migrated images shown in Figures~\ref{fig:marm_duals_WoFS}(e)--(h).
Figure~\ref{fig:marm_duals_WFS} shows the estimated dual data and the corresponding migrated images when free-surface modeling is used.

%%--------------------------------------------------------------------------

%%--------------------------------------------------------------------------

Finally, to assess robustness against kinematic inaccuracies in the background model, we repeated the previous experiment using background model 2, obtained with a 240~m Gaussian window (Figure~\ref{fig:marm_m0}(b)). The acquisition geometry, source wavelet, number of iterations, and the two scenarios---removing or retaining free-surface multiples---were otherwise unchanged.

\begin{figure}[htbp]
    \centering
    \includegraphics[width=0.9\linewidth]{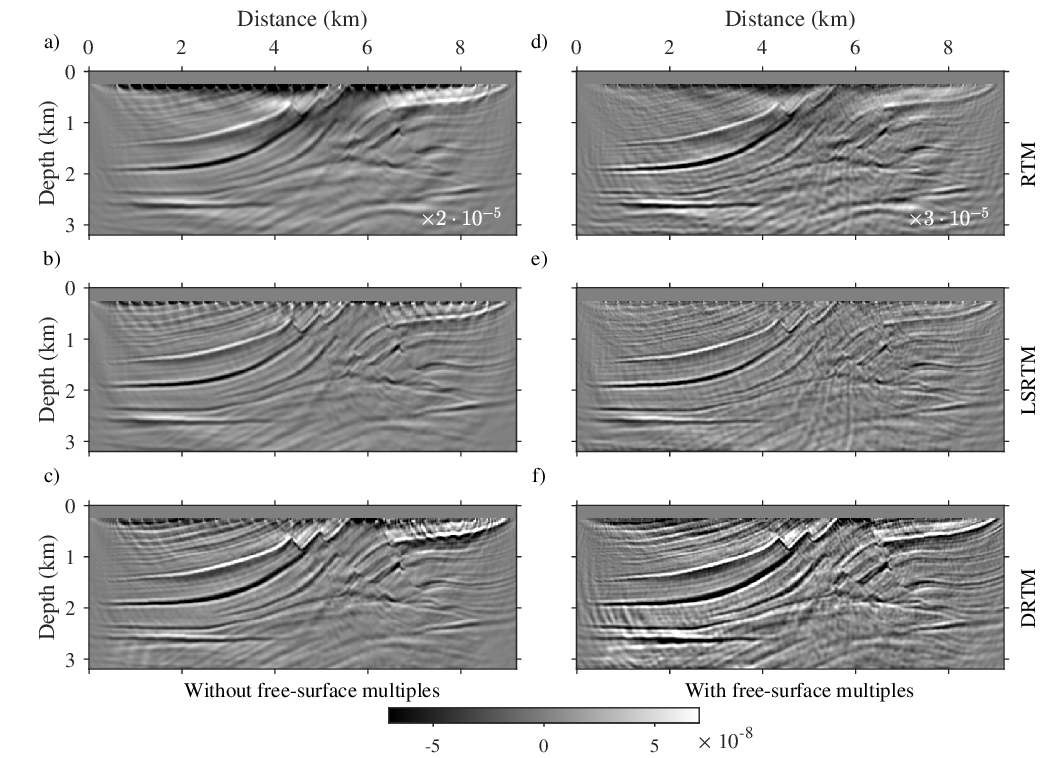}
    \caption{Same as Figure~\ref{fig:marm_results}, but using background model 2 (Figure~\ref{fig:marm_m0}(b)): (a–-c) RTM, LSRTM, and DRTM without free-surface multiples; (d–-f) the same with free-surface multiples.}
    \label{fig:marm_results_240}
\end{figure}

Figures~\ref{fig:marm_results_240}(a)--(c) show the images obtained by RTM, LSRTM, and DRTM, respectively, without free-surface multiples, while Figures~\ref{fig:marm_results_240}(d)--(f) show the corresponding results when free-surface multiples are retained. Despite the lower accuracy of the background model, the DRTM results remain notable. In particular, DRTM retains a clear advantage over RTM and LSRTM in both the shallow and deep parts of the model. The spurious near-vertical events around 5 and 6~km, identified in the RTM/LSRTM images in Figures~\ref{fig:marm_results}(d)--(e), as well as the crosstalk noise, are stronger and more extensive in Figures~\ref{fig:marm_results_240}(d)--(e), as the poorer background model leaves a larger portion of the multiple energy unexplained. This is because the single-scattering nature of RTM and LSRTM, combined with the less accurate background, limits their ability to account for higher-order multiple-scattering waves. In contrast, DRTM converts the multiple energy into improved illumination, as evident in Figure~\ref{fig:marm_results_240}(f), recovering the deep reflectors below~2~km and providing more continuous imaging of the structures beneath the uncovered edges.
Nevertheless, due to the ill-conditioned nature of the inversion problem, some artifacts remain in the DRTM image obtained with background model 2, which is less accurate, such as those in the lower-left corner of Figure~\ref{fig:marm_results_240}(f). Such artifacts can be reduced by applying appropriate regularization in the image space.

%%%%%%%%%%%%%%%%%%%%%%%%%%%%%%%%%%%%%%%%%%%%%%%%%%%%%%%%%%%%%%%%%%%%%%%%%%%%%%%%%%%%%
\subsection{Crustal-scale imaging}
In this section, we assess the performance of DRTM for crustal-scale imaging. As observed in the previous examples, DRTM outperforms LSRTM both computationally and in terms of reconstructed image quality. Therefore, in the following examples, the results are compared only with those obtained using RTM.

\begin{figure}[htbp]
    \centering
    \includegraphics[width=.8\linewidth]{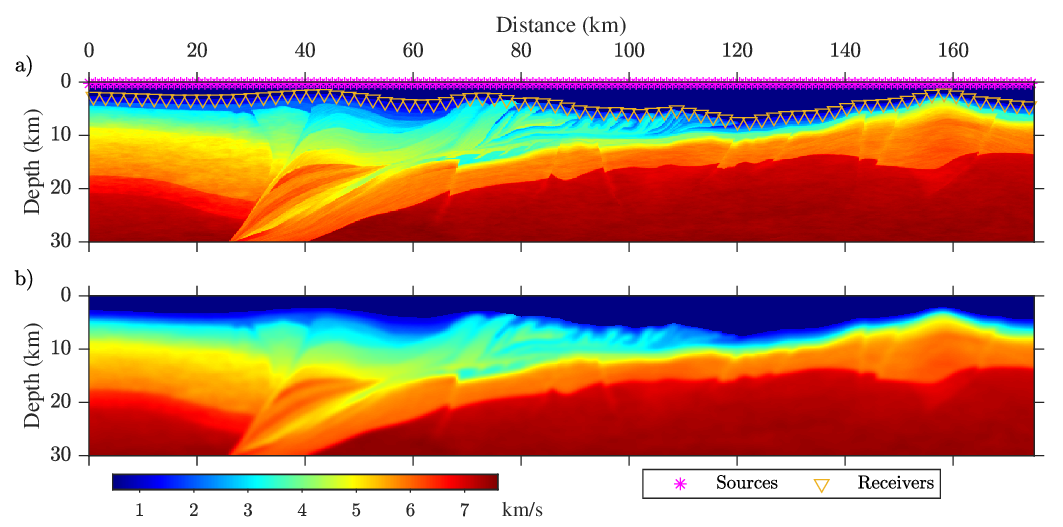}
    \caption{Crustal-scale velocity model. (a) True velocity model with the acquisition geometry overlaid: receivers are shown as orange triangles and sources as purple stars. (b) Migration velocity (background) model obtained by Gaussian smoothing with a window size of 525~m.}
    \label{fig:go3d_models}
\end{figure}

The selected velocity model, the GO\_3D\_OBS model \citep{Gorszczyk_2021_MBG}, represents a complex regional-scale subduction-zone environment spanning 175~km (inline) $\times$ 100~km (crossline) $\times$ 30~km (depth) on a uniform 25~m grid. For this study, we extract a 2D inline section at a crossline distance of 25~km, spanning approximately 175~km in distance and 30~km in depth (Figure~\ref{fig:go3d_models}(a)). The section is resampled to a 75~m spatial interval in both directions, resulting in a computational grid with 401 samples in depth and 2334 in distance. Only the P-wave velocity model is considered, and the density is assumed constant throughout.
The migration velocity model is obtained by applying a Gaussian smoothing operator with a window size of 525~m to the true velocity model (Figure~\ref{fig:go3d_models}(b)).

\begin{figure}[htbp]
    \centering
    \includegraphics[width=0.8\linewidth]{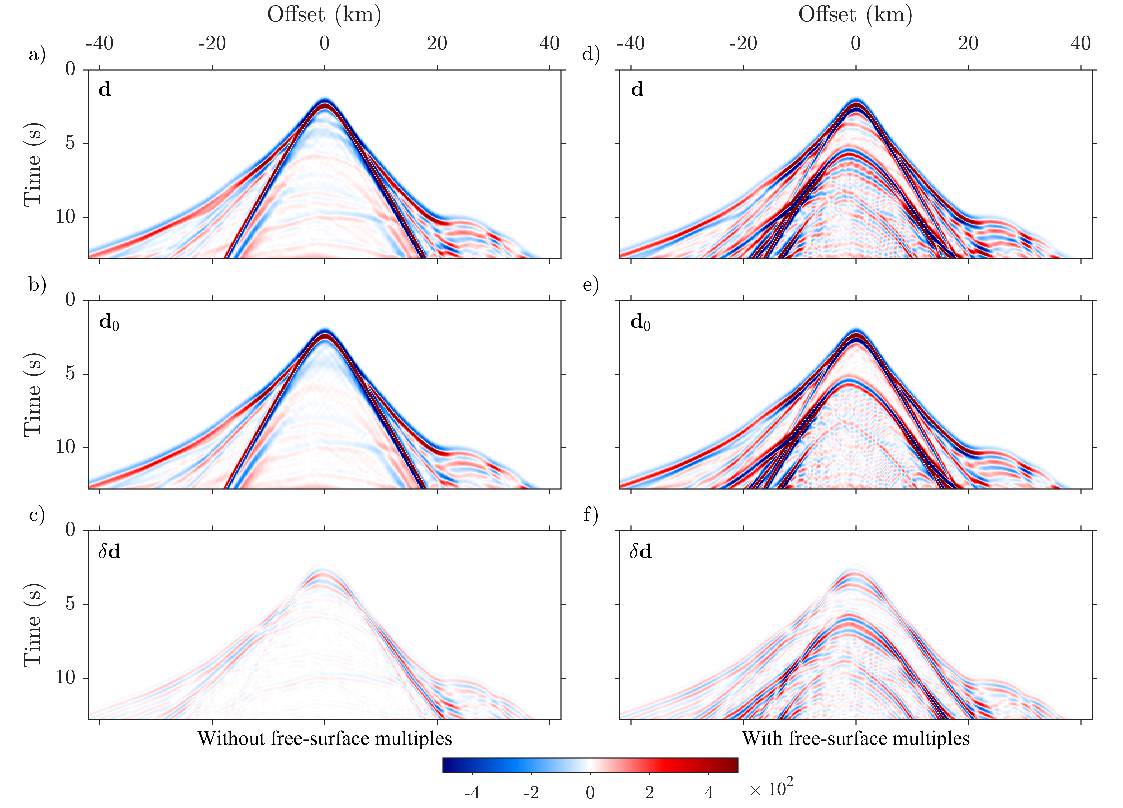}
       \caption{Representative common-shot gathers obtained from the crustal-scale velocity model. (a--b) Data simulated in the (a) true model and (b) background model, without free-surface multiples. (c) Data residual (the difference between (a) and (b)). (d--f) The same as (a--c), respectively, but simulated with free-surface multiples.}
          \label{fig:go3d_data}
\end{figure}

The acquisition consists of 778 sources along the sea surface at 225~m spacing and 78 seafloor receivers at 2250~m intervals. Following seismic reciprocity, source and receiver positions are interchanged during inversion, so that the effective acquisition comprises 78 sources, substantially reducing the number of wave-equation solves. The source is a Ricker wavelet with a peak frequency of 1.3~Hz, and the data are recorded for 14~s with a 4~ms temporal sampling interval. The low source frequency yields a usable signal band extending to approximately 3.5~Hz, consistent with the long-offset, deep-penetration character of the regional acquisition.
Figures~\ref{fig:go3d_data}(a)--(c) show the observed data, the data modeled in the background model, and their difference without free-surface multiples, while Figures~\ref{fig:go3d_data}(d)--(f) show the same obtained by modeling the free-surface multiples.

\begin{figure}[htbp]
    \centering
    \includegraphics[width=.8\linewidth]{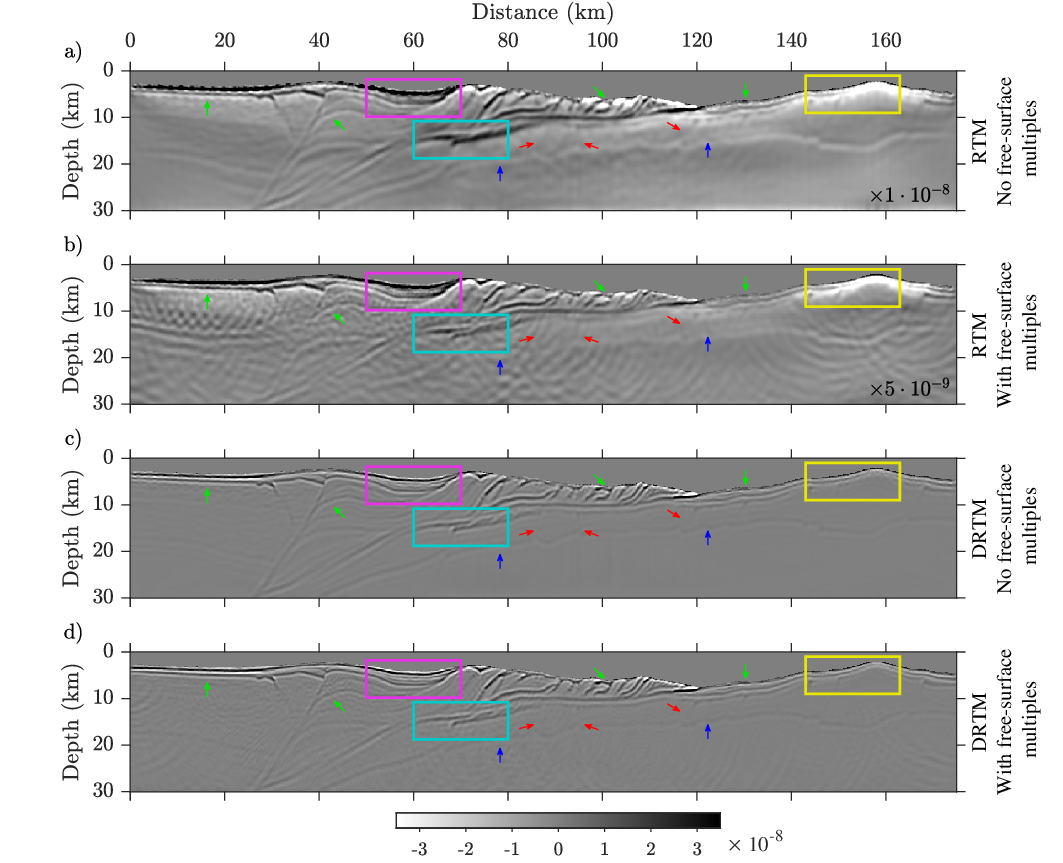}
      \caption{Images obtained from the GO\_3D\_OBS data using (a, b) RTM and (c, d) DRTM, (a, c) without and (b, d) with free-surface multiples. In each panel, the magenta, cyan, and yellow boxes mark the three regions enlarged in Figure~\ref{fig:go3d_results_zoom}, and the red, blue, and green arrows indicate reflectors referred to in the text. All panels share a common color scale.}
    \label{fig:go3d_results}
\end{figure}

\begin{figure}[htbp]
    \centering
    \includegraphics[width=0.45\linewidth]{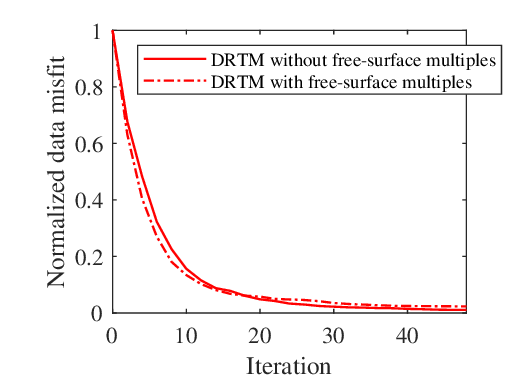}
    \caption{Normalized data misfit versus iteration for DRTM, without and with free-surface multiples.}
    \label{fig:go3d_curve}
\end{figure}

In this experiment, we again investigated imaging of the data while surface-related multiples are considered once as noise and removed (or not simulated) and once as a signal and incorporated into the imaging. Figure~\ref{fig:go3d_results} shows results obtained by RTM (Figure~\ref{fig:go3d_results}(a) and (b)) and DRTM (Figure~\ref{fig:go3d_results}(c) and (d)) using the data without and with surface-related multiples, respectively. We applied 50 iterations to generate the DRTM results while employing the Anderson acceleration with a memory of 6. Figure~\ref{fig:go3d_curve} shows the associated convergence curves. We observe that, even for this large-scale model with a complex wavefield, DRTM successfully fits the full data while keeping the background model fixed.
As in the previous experiments, the RTM amplitudes have been rescaled by a constant factor to share the color scale of the true perturbation and the DRTM image. 
Three sets of colored arrows are drawn at the same positions in every panel of Figure~\ref{fig:go3d_results}: the blue arrows mark the Moho boundary, the red arrows indicate some selected faults, and the green arrows point to near-surface and sedimentary layers. 

We observe that RTM successfully reconstructs interfaces with stronger impedance contrasts (Figures~\ref{fig:go3d_results}(a) and (b)), including the top of the crust between 0 and 40~km, the Moho discontinuity, the dipping structures between approximately 30 and 70~km, and the shallow sedimentary section in the center of the model. However, the results are contaminated by a strong acquisition footprint in the upper few kilometers below the seafloor. The shallow sedimentary interfaces and the top of the crust between 30 and 170~km are not properly recovered. The faulted zones are barely detectable, and the subducting interfaces at depths of around 15~km and distances of 70--90~km are recovered with low resolution.
The DRTM method improved the image by suppressing the shallow acquisition footprints and rebalancing the amplitudes across the entire section (Figures~\ref{fig:go3d_results}(c) and (d)). The shallow sedimentary interfaces, top of the crust, Moho discontinuity, subducting sedimentary interfaces, and fault zones are all reconstructed. The dipping structures and deep reflectors down to approximately 15--20~km are imaged more continuously, while the layered structure on the right side of the model, largely absent from the RTM image, is restored with amplitudes comparable to those in the center. Thus, both the near-surface region beneath the seafloor and the deeper crust benefit from the DRTM method, which compensates for acquisition-footprint effects and illumination decay directly in the data domain rather than through repeated model-space Hessian corrections.

\begin{figure}[htbp]
    \centering
    \includegraphics[width=.8\linewidth]{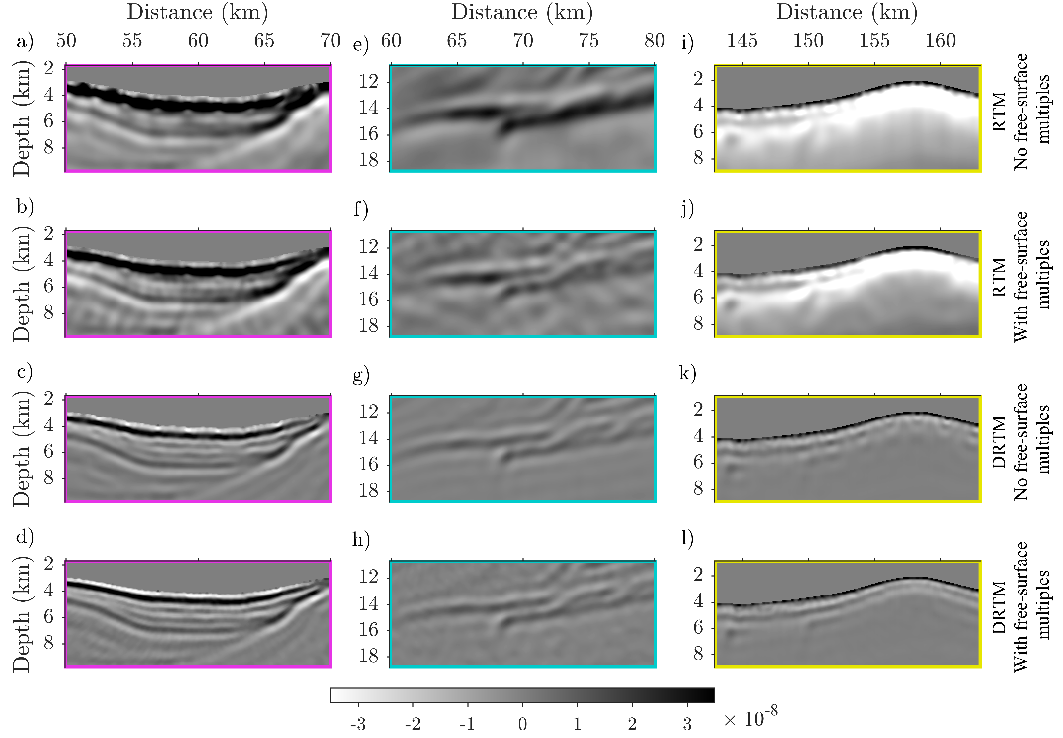}
    \caption{Enlarged views of the three regions marked by the magenta, cyan, and yellow boxes in Figure~\ref{fig:go3d_results}. The columns correspond to the magenta (a--d), cyan (e--h), and yellow (i--l) boxes; within each column, the four rows show, in the same order as Figure~\ref{fig:go3d_results}, RTM without free-surface multiples, RTM with free-surface multiples, DRTM without free-surface multiples, and DRTM with free-surface multiples.}
    \label{fig:go3d_results_zoom}
\end{figure}

The images generated using surface-related multiples (Figures~\ref{fig:go3d_results}(b) and (d)) clearly demonstrate the enhanced illumination provided by the multiples for reconstructing shallow interfaces. Low-frequency noise in the vicinity of the receiver locations is mitigated, and the overall resolution is improved. The regions marked by the colored boxes in Figure~\ref{fig:go3d_results} are enlarged in Figure~\ref{fig:go3d_results_zoom}. Despite the improvement in shallow-interface recovery, Figure~\ref{fig:go3d_results}(b) shows that crosstalk artifacts make the RTM result in the deeper region unreliable. In some locations, these artifacts degrade the images that can be reconstructed using only primary reflections, as shown in Figure~\ref{fig:go3d_results}(a).
The DRTM method, however, can improve the resolution when multiples are included (Figure~\ref{fig:go3d_results}(d) versus Figure~\ref{fig:go3d_results}(c)). Some artifacts remain in the deeper part and along the edges of the model, which could be suppressed by applying an appropriate regularization method.

\begin{figure}[htbp]
    \centering
    \includegraphics[width=1\linewidth]{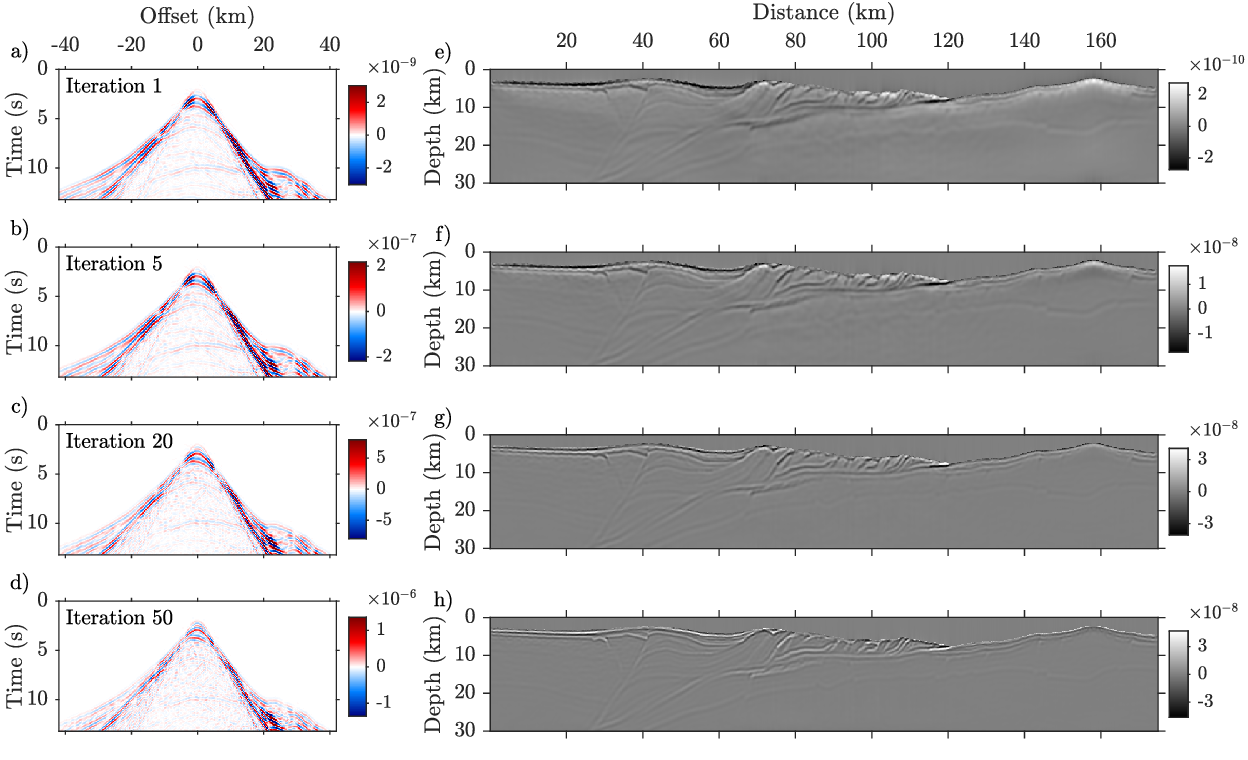}
    \caption{(a--d) Dual data for the case of no free-surface multiples at iterations (a) 1, (b) 5, (c) 20, and (d) 50. (e--h) The migration images obtained from the dual data in (a--d), respectively.}
    \label{fig:go3d_duals_WoFS}
\end{figure}

\begin{figure}[htbp]
    \centering
    \includegraphics[width=1\linewidth]{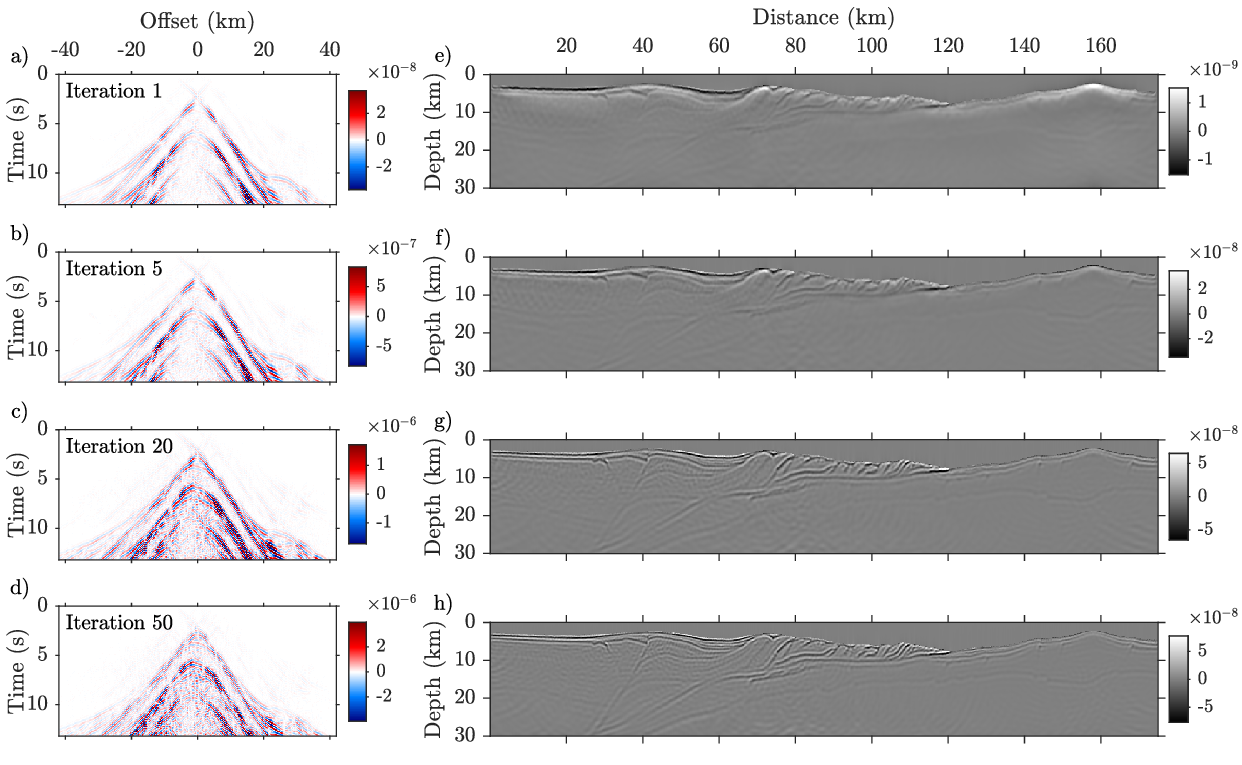}
    \caption{(a--d) Dual data for the case with free-surface multiples at iterations (a) 1, (b) 5, (c) 20, and (d) 50. (e--h) The migration images obtained from the dual data in (a--d), respectively.}
    \label{fig:go3d_duals_WFS}
\end{figure}

The dual data estimated by the DRTM algorithm at different iterations (1, 5, 20, 50) are presented in Figures~\ref{fig:go3d_duals_WoFS}(a)--(d). These dual data contain key information---including all orders of multiple scattering and the effect of the inverse Hessian---required for high-resolution imaging. As the iterations progress, increasingly more information is incorporated into the estimated dual data. This progressive incorporation of information is more clearly visible in the corresponding migrated images shown in Figures~\ref{fig:go3d_duals_WoFS}(e)--(h). Figure~\ref{fig:go3d_duals_WFS} shows the estimated dual data and the corresponding migrated images when free-surface modeling is used.

%%%%%%%%%%%%%%%%%%%%%%%%%%%%%%%%%%%%%%%%%%%%%%%%%%%%%%%%%%%%%%%%%%%%%%%%%%%%%%%%%%%%%
\section{Conclusions} \label{sec:conclusion} 
In this work, we introduced a dual approach to standard Reverse Time Migration (RTM). Unlike standard least-squares RTM, which formulates the inverse problem in the image space and iteratively reconstructs the image by fitting the data using first-order Born modeling, the proposed approach operates in the dual space and moves beyond the first-order approximation. By iteratively updating the dual variables through solutions of the full wave equation, dual RTM reconstructs higher-order scattering and multiple reflections without requiring updates to the background velocity model. From a computational perspective, dual RTM replaces the source-dependent model-space Hessian with a source-independent data-space Hessian. Using a hybrid time--frequency formulation, this Hessian is assembled and inverted once offline in the Fourier domain, reducing the online cost of each dual iteration to that of a single standard RTM application. Synthetic benchmarks across acoustic exploration and crustal-scale models demonstrate that, compared with standard single-scattering migration, the dual formulation can produce higher-quality images with fewer artifacts while requiring less computational effort.

%%%%%%%%%%%%%%%%%%%%%%%%%%%%%%%%%%%%%%%%%%%%%%%%%%%%%%%%%%%%%%%%%%%%%%%%%%%%%%%%%%%%%
\appendix
\section{Construction and direct inversion of the data-space Hessian}
\label{app:hessian}
A central component of both the offline stage and the dual update of Algorithm~\ref{ALG_DRTM} is the construction and inversion of the data-space Hessian $\bQ$ \citep{Gholami_2022_EFW}. In the offline stage, the extended data residual $\delta\bd_s^{e}$ is recovered from the reduced data residual $\delta\bd_s$ by solving the linear system $\bQ\,\delta\bd_s^{e} = \delta\bd_s$, providing the adjoint source from which the forward wavefield is reconstructed (Algorithm~\ref{ALG_DRTM}, Lines~\ref{DRTM_wf_start}--\ref{DRTM_wf_end}). During the dual iterations, the same operator is applied in the dual-update stage (Algorithm~\ref{ALG_DRTM}, Line~\ref{DRTM_dual}) to deblur the data residual evaluated in the perturbed model.

The data-space Hessian $\bQ$ acts on common-source gathers of size $N_r \times N_t \times N_c$ and is therefore a square matrix of order $N_r N_t N_c$, where $N_r$ is the number of receivers, $N_t$ is the number of time samples, and $N_c$ is the number of recorded components in a multicomponent acquisition. Its construction requires $N_r$ PDE solves. However, its explicit construction and inversion in the time domain are computationally prohibitive. Throughout this appendix, we consider a single-component acquisition for simplicity, but the approach can be extended to multicomponent data in the same manner.
While iterative solvers are standard for large linear systems, for the particular problem $\bQ\,\delta\bd_s^{e} = \delta\bd_s$, they require two PDE solves per (inner) iteration, repeated for every source. Since the Hessian must be applied at every dual (outer) iteration, this computational cost makes such methods inefficient and impractical in the present framework.

In this paper, we employ a direct inversion strategy for efficiently inverting $\bQ$ in a hybrid time--frequency domain \citep{Sonbolestan2025}. The data-space Hessian is well approximated by a block-diagonal matrix in the Fourier domain. Each $N_r\times N_r$ block is extracted and inverted independently at each frequency, while wave propagation remains in the time domain. This allows us to compute and store the inverse Hessian only once. It is then used to directly compute $\delta\bd_s^{e} =\bQ^{-1} \delta\bd_s$ in the frequency domain with high accuracy.

Let us define $\delta\hat{\bd}_s(\omega)$ as the frequency slice of a common-source gather associated with the $s$-th source and $\delta\hat{\bd}_s^{e}(\omega)$ the same frequency slice of the shot gather after removing the Hessian effect. Then, for all of the sources $s=1,...,N_s$
\begin{equation} \label{eq:freqsolve}
\delta\hat{\bd}_s^{e}(\omega)
=
\hat{\bQ}(\omega)^{-1}\,\delta\hat{\bd}_s(\omega).
\end{equation}
 Each coefficient matrix $\hat{\bQ}(\omega)$ has size $N_r \times N_r$ ($N_rN_c \times N_rN_c$ for multicomponent data). The details of computing $\hat{\bQ}(\omega)$ are summarized in Algorithm~\ref{ALG2}.

\begin{algorithm}[H]
\caption{Fourier block-diagonal construction and inversion of the data-space Hessian.}
\label{ALG2}
\begin{algorithmic}[1]
\STATE \textbf{Input:} background model $\bm_0$, receiver positions $\bx_r^i$ for $i = 1, \ldots, N_r$, penalty parameter $\mu$.
\STATE Build the weighting function $\mathbf{W}(m,n) = N_t - \max(m,n)$, for $m,n = 0,\dots,N_t-1$
\STATE Compute the receiver-side Green's functions $\mathbf{g}^i(t,\bx)$ for all receivers $i = 1, \ldots, N_r$ by solving the wave equation in background model $\bm_0$ with a Dirac source at each receiver location.
\FOR{$i = 1,\ldots,N_r$}
    \FOR{$j = i,\ldots,N_r$}
        \STATE Form the cross-correlation matrix $\mathbf{G}^{i,j} = \sum_{\bx} \mathbf{g}^i(:, \bx)\, \mathbf{g}^j(:, \bx)^{\top}$.
        \STATE Compute the weighted correlation vector $\mathbf{q}^{i,j}$, by summing the elements of $\mathbf{W} \circ \mathbf{G}^{i,j}$ along its descending diagonals.
        \STATE Compute $\hat{\mathbf{q}}^{i,j}(\omega)$ as the $2N_t$-length FFT of $\mathbf{q}^{i,j}$.
        \STATE For each desired frequency $\omega$, set $\hat{\mathbf{q}}^{i,j}(\omega)$ as the $(i,j)$ entry of $\hat{\bQ}_0(\omega)$, and set $\hat{\mathbf{q}}^{j,i}(\omega) = \text{conj}(\hat{\mathbf{q}}^{i,j}(\omega))$ as the $(j,i)$ entry.
    \ENDFOR
\ENDFOR
\STATE Compute $\hat{\bQ}(\omega)^{-1}=(\hat{\bQ}_0(\omega)+\mu\bI)^{-1}$.
\STATE \textbf{Output:} inverse operators $\hat{\bQ}(\omega)^{-1}$ for desired frequencies.
\end{algorithmic}
\end{algorithm}

%%%%%%%%%%%%%%%%%%%%%%%%%%%%%%%%%%%%%%%%%%%%%%%%%%%%%%%%%%%%%%%%%%%%%%%%%%%%%%%%%%%%%

%  \section*{Acknowledgements}
% This research was funded by the National Science Centre, Poland (NCN), grant nos. 2022/46/E/ST10/00266 and 2023/51/D/ST10/02603. The authors declare no conflicts of interest.

% \section*{Data availability}
% The data and code related to this work are available from the corresponding author upon reasonable request.

% \bibliographystyle{plainnat}
% \bibliography{BIB_arxiv}

\end{document}